\documentclass[pra,aps,twocolumn,showpacs,groupedaddress,superscriptaddress]{revtex4}

	\usepackage{graphicx,epsf,bm,bbm,color,amsmath,ulem,amssymb}

    \newcommand{\s}{\text{St\"uckelberg }}
    \newcommand{\tr}[1]{\textrm{#1}}
	
	\newcommand{\beq}{\begin{equation}}
	\newcommand{\be}{\begin{equation}}
	\newcommand{\beqn}{\begin{eqnarray}}
	\newcommand{\eeq}{\end{equation}}
	\newcommand{\ee}{\end{equation}}
	\newcommand{\eeqn}{\end{eqnarray}}
	\newcommand{\nn}{\nonumber}

\newcommand{\bem}{\begin{pmatrix}}
\newcommand{\eem}{\end{pmatrix}}

\newcommand{\f}{\frac}

\begin{document}

	\today
	
		\title{Geometric phase in St\"uckelberg interferometry}
        \author{Lih-King Lim}
        \affiliation{LCF, Institut d'Optique, CNRS, Univ. Paris-Sud, 2 avenue Augustin Fresnel, F-91127 Palaiseau}
        \affiliation{Max-Planck-Institut f\"ur Physik komplexer Systeme, D-01187 Dresden}
        \author{Jean-No\"el Fuchs}
        \affiliation{LPTMC, CNRS UMR 7600, Universit\'{e} Pierre et Marie Curie, 4 place Jussieu, F-75252 Paris}
        \affiliation{Laboratoire de Physique des Solides, CNRS UMR 8502, Univ. Paris-Sud, F-91405 Orsay}

         \author{Gilles Montambaux}
\affiliation{Laboratoire de Physique des Solides, CNRS UMR 8502, Univ. Paris-Sud, F-91405 Orsay}
\begin{abstract}
We study the time evolution of a two-dimensional quantum particle exhibiting an energy spectrum, made of two bands, with two Dirac cones, as e.g. in the band structure of a honeycomb lattice. A force is applied such that the particle experiences two Landau-Zener transitions in succession. The adiabatic evolution between the two transitions leads to \s interferences, due to two possible trajectories in energy space. In addition to well-known dynamical and Stokes phases, the interference pattern reveals a geometric phase which depends on the chirality (winding number) and the mass sign associated to each Dirac cone, as well as on the type of trajectory (parallel or diagonal with respect to the two cones) in parameter space. This geometric phase reveals the coupling between the bands encoded in the structure of the wavefunctions. 
\end{abstract}
	\maketitle

\section{Introduction}
\s interferometry is the realization of an interferometer for a quantum particle with an energy spectrum possessing at least two branches, or bands, separated by a gap (for example, due to a band structure). The problem was originally raised in the context of slow atomic/molecular collisions experiencing multiple electronic transitions \cite{S1932}, where each transition is modeled by the Landau-Zener (LZ) tunneling process \cite{LZ1932}. It has since been mapped onto a wide class of systems described by a two-level time-dependent Hamiltonian with multiple avoided crossings, including the microwave excitation of Rydberg atoms \cite{Baruch1992,Yoakum1992}, superconducting qubits \cite{SAN2010,Pekola2011,Huang2011}, quantum wires \cite{DF2013}, as well as Bose-Einstein condensates in optical lattices \cite{Kling2010,Zenesini2010}.
For a general review of St\"uckelberg interferometry, see Ref. \cite{SAN2010}.
\begin{figure}[ht!]
\begin{center}
\includegraphics[width=8cm]{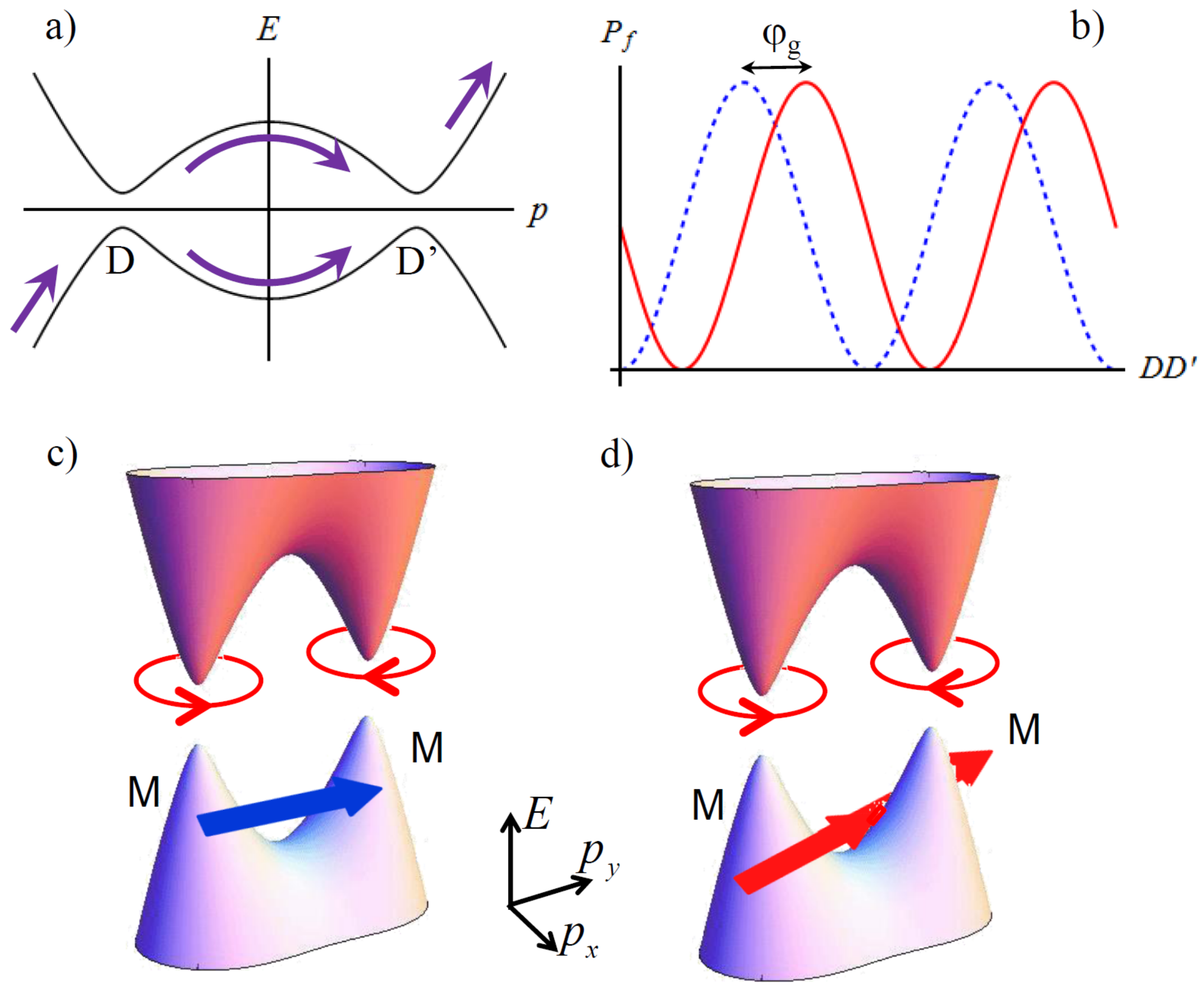}
\end{center}
\caption{a) A St\"uckelberg interferometer made of two avoided crossings ($D$ and $D'$) in an energy spectrum $E$ as a function of momentum $p$. A particle initially in the lower band is forced through the two avoided crossings that act as beam splitters. $P_f$ is the probability for the particle to end up in the upper band. This can occur through two different paths in energy-momentum space. b) The geometric phase $\varphi_g$ is revealed in the interference pattern ($P_f$ as a function of the distance $DD'$). Dashed blue line ($\varphi_g=0$) corresponds to trajectory (c) and full red line ($\varphi_g\neq 0$) to trajectory (d). c) and d) Double Dirac cone energy spectrum as a function of two-dimensional momentum with parallel (c, blue arrow) or diagonal (d, red arrow) trajectories. Chirality (shown as directed circle) and mass $M$ of each Dirac cone are also indicated (see text).}
\label{stuckinter}
\end{figure}

Recently, topological bandstructure engineering has attracted a lot of interest both in condensed matter systems \cite{Hasan2010,Qi2011} as well as in artificial crystals \cite{Polini2013} simulated by various means such as cold atoms in optical lattices \cite{Soltan2011,Tarruell2012,Atala2013,Jotzu2014,Aidelsburger2014,Duca2014}, microwave resonators \cite{Bellec2013} or polaritons \cite{Jacqmin2014}. In these systems, Dirac cones in the Bloch energy spectrum are the basic entity of interest \cite{Neto2009,AsanoHotta2012}. Moreover, the construction of topological bands can be induced by a modification of the local character of Dirac cones, e.g., by changing the relative signature of the masses of two Dirac cones \cite{Haldane1988}. While the hallmark of a simple topological state is displaying perfectly quantized conductance at the edge (or boundary) \cite{Hasan2010}, it is interesting to search for measurable bulk topological signatures in these new systems \cite{Xiao2010,Nagaosa2010,Jotzu2014,Gorbachev2014}.

In this work, we consider a \s interferometer made of two massive Dirac cones in two dimensions (2D). By accelerating a quantum particle through the two cones in succession, non-adiabatic processes at the two avoided crossings (described by Landau-Zener tunnelings) coherently split and recombine the wavefunction, see Fig. \ref{stuckinter}a. The final transition probability oscillates in magnitude due to interferences between the two possible paths in the energy space, as the phase accumulated along the path is varied (Fig. \ref{stuckinter}b). An analogy can be drawn with the optical Mach-Zehnder interferometer, except that with a \s interferometer the motion of the quantum particle takes place in the energy-momentum plane instead of the real space $x$-$y$ plane. The avoided crossings play the role of the optical beam splitters and the two adiabatic energy bands (upper and lower bands) are the analogue of the two optical arms \cite{Zeilinger1981,Holbrow2002}. It is also important to realize that the \s interferometer here deals with spinorial and not scalar waves. The internal degree of freedom is related to the band index (lower or upper band) which arises, for example, from the pseudospin-$1/2$ sublattice degree-of-freedom in a honeycomb tight-binding system. In an optical Mach-Zehnder interferometer, this role would be played by the polarization of light. As anticipated long ago by Pancharatnam \cite{Pan1956}, the phase and the contrast of interferences can be modified by the polarization degree of freedom. The purpose of the present article is to study the influence of the pseudospin degree of freedom of the quantum particle on the \s interferometer.

We show that, in addition to the well-known dynamical phase which depends on the energy separation between the two bands and the Stokes phase accumulated at the LZ transitions, there is a geometric contribution which has the form of a gauge-invariant open-path geometric phase (also called noncyclic geometric phase) \cite{SB1988,PS1998}. The later being a generalization of the well-known Berry phase \cite{Berry1984}. This is the central result of this work, as first anticipated by us in a recent letter \cite{LFM2014}. This geometric phase depends on the chirality and the mass of the Dirac cones, as well as the type of trajectory crossing the two Dirac cones. As an illustration, Fig. \ref{stuckinter}c and d show two different trajectories for a double cone energy spectrum with given  masses and chiralities. They both correspond to the same energy landscape (Fig. \ref{stuckinter}a) but result in different interference pattern, see the two curves Fig. \ref{stuckinter}b.

We stress the differences with recent interferometric studies with Dirac cones where only adiabatic evolution within a \textit{single band} is considered \cite{Abanin2013,Liu2013,Duca2014}. Here, non-adiabatic transitions between \textit{two bands} are required to realize the \s interferometer. Moreover, unlike the single avoided crossing problem (a Landau-Zener problem) an exact solution to the problem of a double Landau-Zener Hamiltonian does not generally exist \cite{Shimshoni1991,Suominen1992,FLM2012}. The theoretical framework we employ is therefore founded on two approximation schemes, i.e., the so-called adiabatic impulse model \cite{SAN2010}, where the two Landau-Zener tunneling events are taken to be independent, and the adiabatic perturbation theory.

The paper is organized as follows. In section \ref{sec:stuckhamil} we introduce four classes of Bloch Hamiltonians featuring a pair of Dirac cones. Then by considering two types of trajectories in the parameter space, we obtain eight time-dependent Hamiltonians for St\"uckelberg interferometry. In section \ref{sec:stucktheo}, we provide a heuristic but general solution to the interferometer problem based on \s theory and showing the presence of a non-trivial geometric phase affecting the interference pattern. In section \ref{sec:statement}, we mathematically formulate the dynamics of a quantum particle going through such an interferometer. In sections \ref{sec:twobands}--\ref{sec:geointer}, we consider the specific case of a double cone with the same mass, opposite chirality and a diagonal trajectory and contrast it with that of a parallel trajectory studied in Ref. \cite{FLM2012}. In section \ref{sec:twobands}, we first show numerically the presence of a phase shift. Then we compute the geometric phase using different basis and gauge choices. In section \ref{sec:adiatheo}, we give its analytic derivation using adiabatic perturbation theory. We then study the special massless limit in section \ref{sec:massless}. Section \ref{sec:geointer} provides a geometrical interpretation of the geometric phase on the Bloch sphere. In section \ref{sec:map} we give the geometric phase for the eight types of \s interferometers. And we conclude in section \ref{sec:conc}.

\section{Models and Statement of the Problem}\label{sec:stuckhamil}
A Dirac cone in the energy spectrum displays interesting topological character related to the (pseudo-)spinorial nature of the associated wavefunction. To give an example that highlights the importance of the pseudospin structure, it essentially determines the Chern number of a 2D energy band in the modern topological characterization of bandstructure \cite{Hasan2010}. To reveal this pseudospin structure in \s interferometry, we consider the low-energy description of a given pair of inequivalent Dirac cones, inspired by the merging transition of Dirac points in uniaxially deformed graphene \cite{Montambaux2009,Tarruell2012,Lim2012}.

\subsection{Four classes of Bloch Hamilonians featuring a pair of Dirac cones}
By restricting to Dirac cones with $\pm 1$ topological charges (see below), we begin by introducing two broad classes \cite{Montambaux2009,Lim2012,deGail2011} of Bloch Hamiltonians \cite{footnoteblochhamil}:
\subsubsection{Dirac cone pair with opposite chirality}
The first class is given by the low energy expansion
\beq\label{uni}
H(\vec{p})=\biggl(\f{p_x^2}{2m}-\Delta_* \biggr)\sigma_x+c_y p_y \sigma_y+ M_z(\vec{p})\sigma_z,
\eeq
where $\vec{p}=(p_x,p_y)$ is the long wavelength quasimomentum (a parameter, not an operator), $m$ gives the band curvature in the $x$-direction and $c_y>0$ is the $y$-direction velocity. The Pauli matrices $\sigma_{x,y,z}$ operate in the pseudospin space, which stems from a sublattice degree of freedom of the microscopic 2D tight-binding lattice model of graphene. In other words, the Hamiltonian is the low-energy Bloch Hamiltonian centered at the midpoint in reciprocal space between the two Dirac cones. The function $M_z(\vec{p})$ opens a gap at the two Dirac cones and is usually referred to as a ``mass''. We will consider two such mass functions: either $M_z(\vec{p})=M$ is a constant or $M_z(\vec{p})=c_xp_x$ changes sign between $p_x<0$ and $p_x>0$ assuming that the velocity parameter $c_x>0$ (see end of the section for their physical meanings).

\begin{figure}
\begin{center}
\includegraphics[width=6.6cm]{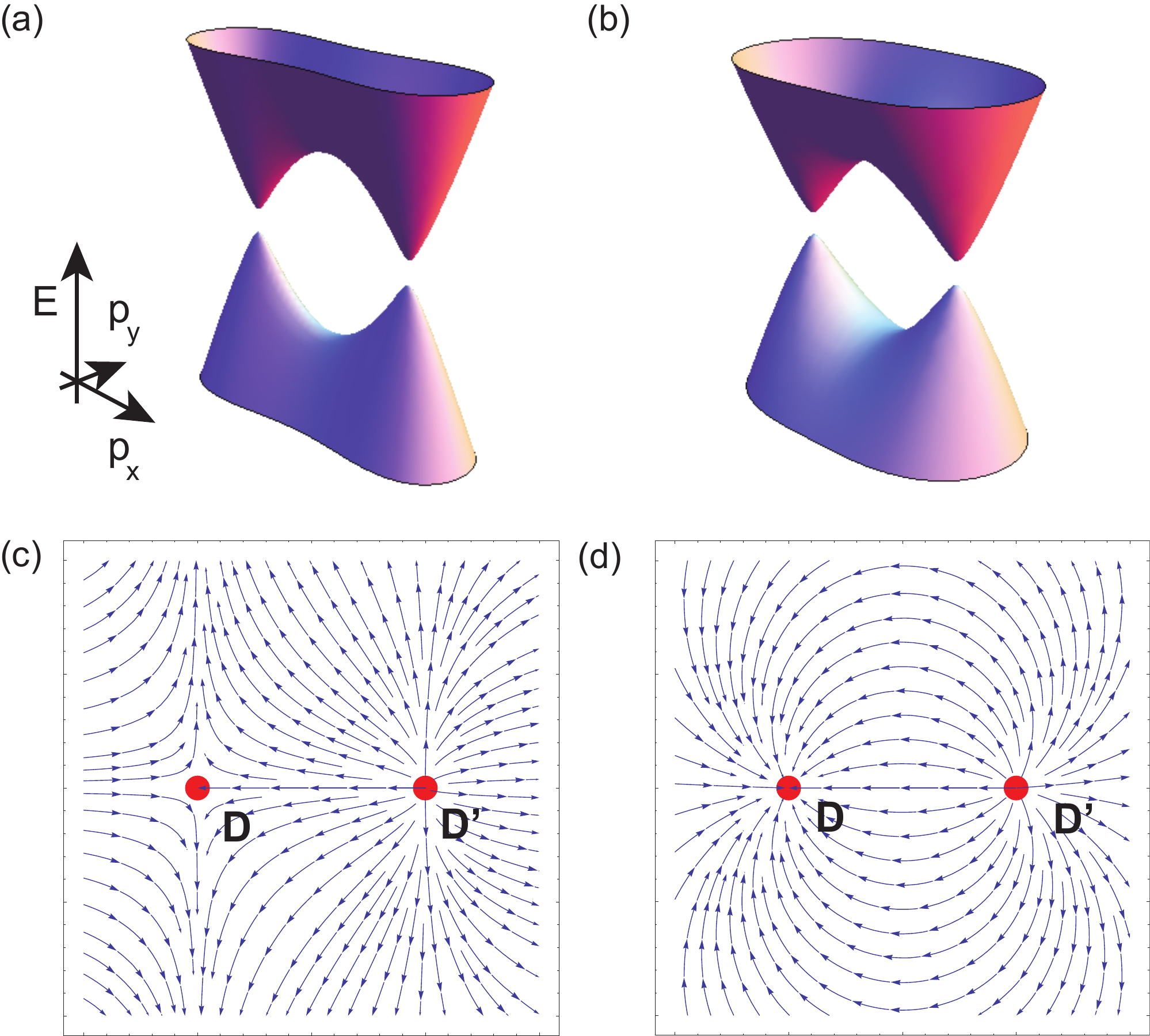}
\end{center}
\caption{(a,b) Low energy spectra featuring two gapped Dirac cones: (a) corresponds to the universal model with opposite chiralities, Eq.(\ref{uni}); (b) corresponds to the universal model with identical chiralities, Eq.(\ref{uni2}). (c,d) Plot of the relative phase $\phi$ between $\sigma_x$ and $\sigma_y$ components of the Hamiltonian (azimuthal phase on the Bloch sphere) as a function of the momentum $(p_x,p_y)$, with Dirac points located at $D,D'$. (c) Hamiltonian with opposite chirality (winding number $-1$ at $D$ and $+1$ at $D'$) (d) Hamiltonian with same chirality (winding number $+1$ at $D$ and $D'$).}
\label{vortices}
\end{figure}

The properties of this first class of Bloch Hamiltonians are, firstly, that the energy spectrum is:
\beq
E_\pm(\vec{p})=\pm\biggl[\bigl(\f{p_x^2}{2m}-\Delta_* \bigr)^2+ c_y^2 p_y^2+M_z(\vec{p})^2   \biggr]^{1/2}\, .
\eeq
The two gapped Dirac cones lie on the $p_y=0$ axis and $\Delta_*\geq 0$ determines the distance between the two cones located at valleys $\vec{p}=D,D'\approx (\mp\sqrt{2 m\Delta_*},0)$, see Fig. \ref{vortices}. The gap is $2|M_z(D,D')|$.

\begin{figure}
\begin{center}
\includegraphics[width=4.cm]{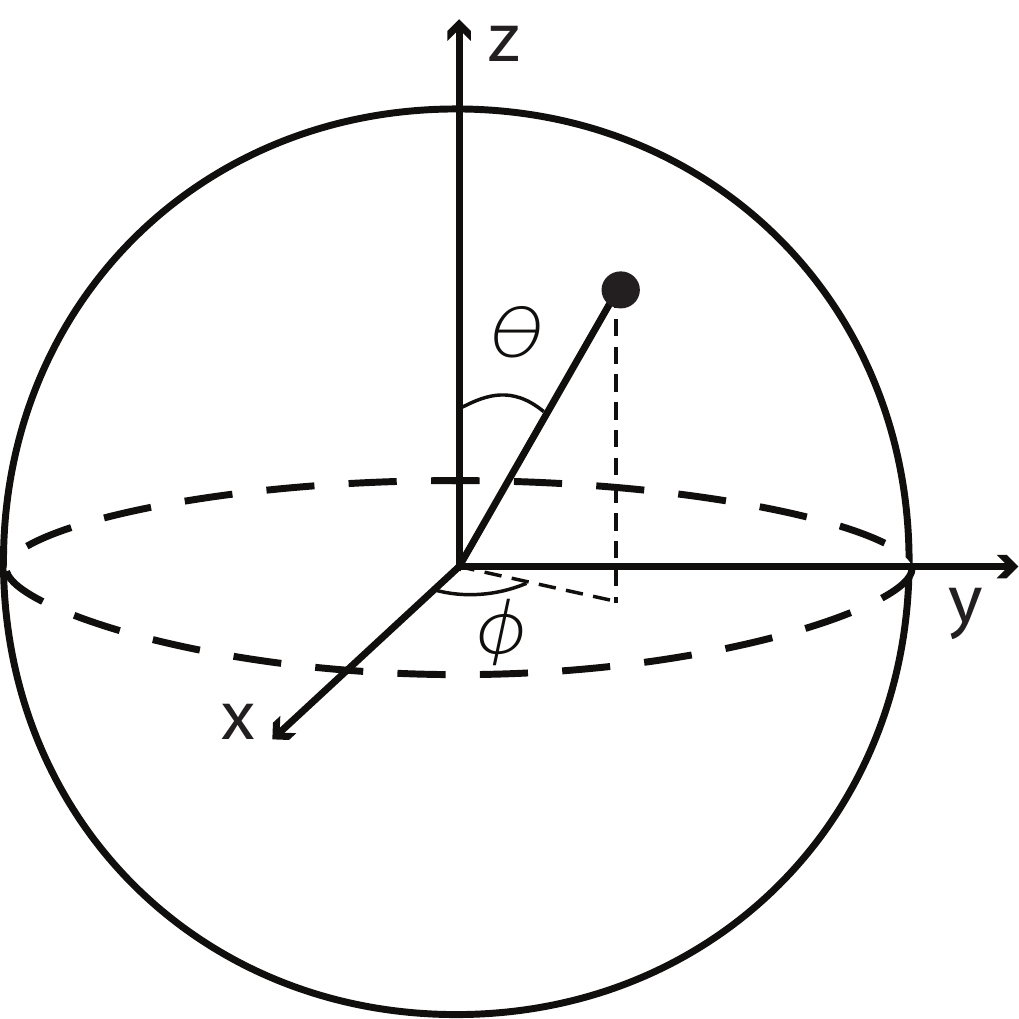}
\end{center}
\caption{Bloch sphere representation of a Zeeman-like Hamiltonian $H=\vec{B}\cdot \vec{\sigma} = E_+ \vec{n}\cdot \vec{\sigma}$ where the unit vector $\vec{n}$ is parameterized by spherical coordinates given by the polar angle $0\leq\theta\leq\pi$ and the azimuthal angle $0\leq\phi<2\pi$.}\label{fig:blochh}
\end{figure}
Secondly, the Dirac cones are characterized by their chirality, or winding number. This is a property of the eigenstates $|\psi_{\pm}(\vec{p})\rangle$ or of the Bloch Hamiltonian $H(\vec{p})$ that is not apparent in the energy spectrum. In order to reveal it, we parameterize the $2\times 2$ Bloch Hamiltonian (\ref{uni}) as a Zeeman-like Hamiltonian for a spin $\vec{\sigma}$ in a magnetic field $\vec{B}(\vec{p})$ such that
\beq
H(\vec{p})=\vec{B}(\vec{p})\cdot \vec{\sigma} = E_+(\vec{p})\,\vec{n}(\vec{p})\cdot \vec{\sigma}
\eeq
where $\vec{n}(\vec{p})$ is a 3D unit vector living on a Bloch sphere $S^2$ (see Fig.~\ref{fig:blochh}). In spherical coordinates $\vec{n}=[\sin\theta \cos\phi,\sin\theta \sin \phi,\cos\theta]$, where $\theta$ is the polar angle (from the north pole between $0$ and $\pi$) and $\phi$ is the azimuthal angle (along the equator from $0$ to $2\pi$). At each point in the 2D quasimomentum $\vec{p}=(p_x,p_y)$ space, we associate the unit vector $\vec{n}(\vec{p})$ that gives rise to the pseudospin texture mentioned at the beginning of this section. One interesting quantity to examine is the azimuthal angle $\phi$ as a function of $\vec{p}$, see Fig.~\ref{vortices}, where we notice the presence of quantized vortices located at the position of the Dirac cones in the energy spectrum, i.e. $\vec{p}=D,D'\approx (\mp\sqrt{2 m\Delta_*},0)$. Note that the existence of these vortices is independent of $M_z(\vec{p})$ being zero or not, i.e. it is not tied to the existence of contact points (Dirac points) in the energy spectrum. These vortices carry opposite topological charges $W=\pm 1$, known as a chirality or winding number, see Fig. \ref{vortices}(a). This can be computed on a line integral on a contour encircling $D$ or $D'$ as $W=(1/2\pi)\oint d \vec{k}\cdot \vec{\nabla}_{\vec{k}}\phi$. When $M_z=0$, the two Dirac cones are gapless. In that case, the Berry phase acquired when encircling a single Dirac cone is quantized to $\pm\pi$ (note that the Berry phase is defined modulo $2 \pi$). This is no longer true upon opening a gap $M_z\neq 0$, although the quantized vortices are still present (see, e.g., the discussion of that point in Ref.~\cite{Fuchs2010}). We thus see that the opening of an energy gap, even though rendering the bandstructure semiconductor-like, merely modifies the orientation of the pseudospin direction while band coupling effects remain important. As we already mentioned, for a gapped spectrum both the signs of the ``masses" $\tr{sgn}[M_z(D,D')]$ and their chirality are relevant information for determining the Chern number of that band, see e.g., Ref. \cite{Hasan2010,Sticlet2012}.

\medskip
\subsubsection{Dirac cone pair with same chirality}
The second class of Bloch Hamiltonians is given by \cite{deGail2011}
\beq\label{uni2}
H(\vec{p})=\biggl(\f{p_x^2-p_y^2}{2m}-\Delta_* \biggr)\sigma_x+\f{p_x\,p_y}{m} \sigma_y+ M_z(\vec{p})\sigma_z.
\eeq
In this case, the energy spectrum $E_\pm (\vec{p})=\pm \sqrt{\left(\f{p_x^2-p_y^2}{2m}-\Delta_* \right)^2+\left(\f{p_x\,p_y}{m}\right)^2+M_z(\vec{p})^2}$ is qualitatively similar to the previous case, see Fig. \ref{vortices}, featuring two gapped Dirac cones at $\vec{p}=(D,D')\approx (\mp \sqrt{2m\Delta_*},0)$. The crucial difference is that here, the two Dirac cones possess the \textit{same} chirality. This is most clearly seen by plotting the corresponding azimuthal phase $\phi(\vec{p})$, see Fig. \ref{vortices}(b). The two vortices with topological charge $+1$ are clearly seen. For this Bloch Hamiltonian, we will also consider two different mass functions $M_z(\vec{p})=M$ or $c_xp_x$.

\medskip

\subsubsection{Physical examples}
In order to refer to these four cases, we introduce the following notations.
Let $\chi$ be the product of the chirality of the two cones ($\chi=\pm1$), and $\mu$ be the product of the mass sign of the two cones ($\mu=\pm 1$). The four classes of Bloch Hamiltonians parameterized by $(\chi,\mu)=(\pm,\pm)$ becomes
\beqn
H_{\chi,\mu}(\vec{p})=X_\chi(\vec{p})\,\sigma_x+Y_\chi\,(\vec{p})\sigma_y+Z_\mu(\vec{p})\,\sigma_z,
\label{4h}
\eeqn
with $X_\chi(\vec{p})$, $Y_\chi\,(\vec{p})$ and $Z_\mu(\vec{p})$ summarized in Table \ref{table2}.
\begin{table}
\caption{Summary of four classes of Bloch Hamiltonians $H_{\chi,\mu}(\vec{p})=X_\chi(\vec{p})\,\sigma_x+Y_\chi\,(\vec{p})\sigma_y+Z_\mu(\vec{p})\,\sigma_z$ with $(\chi,\mu)=(\pm,\pm)$ depending on the chirality product $\chi$ and the mass sign product $\mu$.}\label{table2}
\vspace{0.25cm}
\begin{tabular}{c|c|c|c}
($\chi$,$\mu$)  &  $X_\chi(\vec{p})$ & $Y_\chi(\vec{p})$ & $Z_\mu(\vec{p})=M_z(\vec{p})$  \\
\hline
($-$, $+$)& $\f{p_x^2}{2m}-\Delta_*$ & $c_y p_y$ & $M$\\
($-$, $-$) & $\f{p_x^2}{2m}-\Delta_*$ & $c_y p_y$ & $c_x p_x$\\
($+$, $+$) & $\f{p_x^2-p_y^2}{2m}-\Delta_*$ & $\f{p_x p_y}{m}$ & $M$\\
($+$, $-$) & $\f{p_x^2-p_y^2}{2m}-\Delta_*$ & $\f{p_x p_y}{m}$ & $c_x p_x$\\
\end{tabular}
\end{table}
The physical meaning of the four Bloch Hamiltonians becomes clear. For $(\chi,\mu)=(-,+)$, it describes a pair of Dirac cones with opposite chirality and a constant mass function. This is the low-energy Hamiltonian describing gapped graphene due to inversion symmetry breaking (as boron nitride, e.g.) \cite{Semenoff1984}. For $(\chi,\mu)=(-,-)$, it corresponds to a pair of Dirac cones with opposite chirality but with a momentum-dependent mass function such that it gives an opposite sign in between the two valleys. This describes the case of a Chern insulator as, e.g. the Haldane model in the non-trivial phase \cite{Haldane1988}. Thirdly with $(\chi,\mu)=(+,+)$, it corresponds to a pair of Dirac cones having the same chirality and a constant mass function. This is the case of a twisted graphene bilayer in which each of the two quadratic band contact points of the untwisted bilayer splits in two linear band contact points (i.e. Dirac points) with identical chirality \cite{deGail2011}. And finally with $(\chi,\mu)=(+,-)$, it is a pair of Dirac cones with the same chirality and a momentum-dependent mass function.

\subsection{Eight time-dependent Hamiltonians for St\"uckelberg interferometry}\label{sec:8cases}
To realize a St\"uckelberg interferometer with $H_{\chi,\mu}(\vec{p})$, we now subject the particle to a constant force $\vec{F}$ so as to drive the particle through two avoided crossings in the vicinity of the two Dirac cones. The phenomenon is equivalent to realizing Bloch oscillations by subjecting a Bloch electron to a constant electric field. The constant force can be implemented by using a time-dependent gauge potential while preserving the crystal symmetry of the lattice. This permits the description of Bloch Hamiltonian, albeit with the modification that the gauge-invariant quasimomentum is now given by the sum of the original quasimomentum (without the external field) and a time-dependent uniform vector potential $\vec{p}+\vec{F}t$ thus rendering the Bloch Hamiltonian time-dependent $H(\vec{p})\rightarrow H(\vec{p}+\vec{F}t)$ \cite{Xiao2010,FLM2012,footnoteblochhamil}. An equivalent viewpoint is to implement the force directly as a spatial potential with a constant gradient, which results in the same time-dependent Bloch Hamiltonian, see appendix~\ref{sec:altder}.

\begin{figure}
\begin{center}
\includegraphics[width=8.5cm]{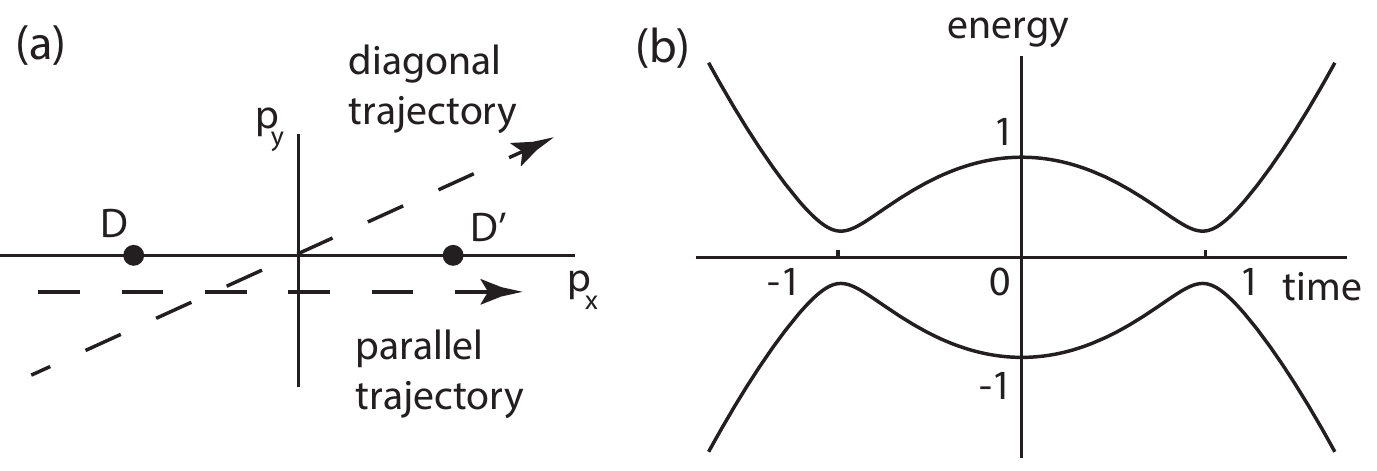}
\end{center}
\caption{(a) Two trajectories (parallel or diagonal) realized by applying a constant force $\vec{F}$. (b) Energy landscape as seen by the particle under acceleration (similar for both trajectories).}
\label{traj}
\end{figure}

Given the two Dirac points $D,D'$ of interest, we consider two types of straight trajectories in the quasimomentum space governed by the direction of $\vec{F}$, see Fig. \ref{traj}a.
We introduce a new index $\tau$ for the two trajectories: $\tau=+1$ for a parallel trajectory and $\tau=-1$ for a diagonal trajectory, respectively. When $\tau=+1$, we substitute $(p_x,p_y)\rightarrow (F_x t,p_y)$ in $H_{\chi,\mu}(\vec{p})$. The trajectory is parallel with the axis of the Dirac cones, and the ``distance" from that axis is set by the constant $p_y$. For $\tau=-1$, we substitute $(p_x,p_y)\rightarrow (F_x t,F_y t)$ in  $H_{\chi,\mu}(\vec{p})$. The diagonal trajectory crosses the midpoint of the line connecting the two Dirac cones (in this case the $x$-axis), with an angle $\arctan (F_y/F_x)$. We finally arrive at the eight time-dependent Hamiltonians ($(\chi,\mu,\tau)=(\pm,\pm,\pm)$) for the \s interferometer given by
\beqn
H_{\chi,\mu,\tau}(t)=X_{\chi,\tau}(t)\,\sigma_x+Y_{\chi,\tau}\,(t)\sigma_y+Z_{\mu}(t)\,\sigma_z
\eeqn
with the components summarized in Table \ref{table3}. The adiabatic energy spectrum takes the form $E_\pm(\vec{p})\rightarrow E_\pm(t)$ featuring two avoided crossings as expected, see Fig. \ref{traj}b.

\subsection{Statement of the problem}

For the $2\times 2$ time-dependent Hamiltonian $H_{\chi,\mu,\tau}(t)$ of the above form, the state $|\psi(t)\rangle$ evolves according to the time-dependent Schr\"{o}dinger equation $i\frac{d}{dt} |\psi(t)\rangle = H_{\chi,\mu,\tau}(t)|\psi(t)\rangle$. The instantaneous eigenstates corresponding to upper (lower) energy bands are defined in the usual way $H_{\chi,\mu,\tau}(t)|\psi_\pm (t)\rangle = E_\pm(t) |\psi_\pm (t)\rangle$. We note that the eigenstates are spinors (they can be represented on a Bloch sphere) and are defined up to a gauge choice (i.e. a time-dependent phase choice). The \s interferometer problem is to compute the transition probability $P_f=|\langle \psi_+(+\infty)|\psi(t\to +\infty)\rangle|^2$, namely the probability for a particle to end up in the upper band in the far future $|\psi_+(+\infty)\rangle$ given the initial state (in the far past) in the lower band $|\psi(t\to -\infty)\rangle\equiv |\psi_-(-\infty)\rangle$.

\section{St\"uckelberg theory including a geometric phase}\label{sec:stucktheo}
We first give a heuristic solution of the interferometer problem, based on \s theory. The spirit is to treat each avoided crossing as an independent Landau-Zener tunneling event and paying extra attention to the adiabatically accumulated phase in between the two LZ events. Everything non-adiabatic is assumed to occur at the LZ events.

\begin{table}
\caption{Eight $2\times 2$ time-dependent Hamiltonians parametrized by the chirality product $\chi$, the mass sign product $\mu$ and the trajectory type $\tau$: $H_{\chi,\mu,\tau}(t)=X_{\chi,\tau}(t)\,\sigma_x+Y_{\chi,\tau}\,(t)\sigma_y+Z_{\mu}(t)\,\sigma_z $ with $(\chi,\mu,\tau)=(\pm,\pm,\pm)$.}\label{table3}
\vspace{0.25cm}
\begin{tabular}{c|c|c|c|c}
\#&$(\chi,\mu,\tau)$ &$X_{\chi,\tau}(t)$ &  $Y_{\chi,\tau}\,(t)$ & $Z_{\mu}(t)$ \\
\hline
1&($-$, $+$, $+$) & $\f{F_x^2t^2}{2m}-\Delta_*$ & $c_y p_y$ & $M$\\
2&($-$, $+$, $-$) & $\f{F_x^2t^2}{2m}-\Delta_*$ & $c_y F_y t$ & $M$ \\
3&($-$, $-$, $+$) & $\f{F_x^2t^2}{2m}-\Delta_*$ & $c_y p_y$ & $c_x F_x t$  \\
4&($-$, $-$, $-$) & $\f{F_x^2t^2}{2m}-\Delta_*$ & $c_yF_yt$ & $c_x F_x t$\\
5&($+$, $+$, $+$) & $\f{F_x^2t^2}{2m}-\f{p_y^2}{2m}-\Delta_*$ & $\f{F_x p_y t }{m}$ & $M$  \\
6&($+$, $+$, $-$) & $\f{F_x^2t^2}{2m}-\f{F_y^2t^2}{2m}-\Delta_*$ & $\f{F_x F_y t^2}{m}$ & $M$ \\
7&($+$, $-$, $+$) & $\f{F_x^2t^2}{2m}-\f{p_y^2}{2m}-\Delta_*$ & $\f{F_x p_y t}{m}$ & $c_x F_x t$ \\
8&($+$, $-$, $-$) & $\f{F_x^2t^2}{2m}-\f{F_y^2t^2}{2m}-\Delta_*$ & $\f{F_x F_y t^2}{m}$ & $c_x F_x t$ \\
\end{tabular}
\end{table}

For a single linear crossing, usually described by a Hamiltonian $H(t)=At\sigma_z+V\sigma_x$ (with $A$ and $V \in \mathbb{R}$) in the vicinity of the crossing assumed at $t=0$, the Landau-Zener (LZ) tunneling probability is $P_{LZ}=e^{-\pi V^2/(|A| \hbar)}=e^{-2\pi \delta}$  where $\delta=V^2/(2 |A| \hbar) \sim \textrm{gap}^2/(\hbar \cdot \textrm{force} \cdot \textrm{speed})$ is the adiabaticity parameter ($\delta \to \infty$ in the adiabatic limit).

In the \s interferometer problem $H_{\chi,\mu,\tau}(t)$ with two linear avoided crossings, we follow Ref.~\cite{SAN2010} using the so-called adiabatic impulse model, which is valid in the \s regime, i.e., assuming the two LZ events are independent. This means that the time spent between the two avoided crossings should be much larger than the tunneling time, which may be estimated as $\tau_{LZ}\sim \frac{\hbar}{|V|}\textrm{max}(\delta,\sqrt{\delta})$ \cite{SAN2010}. Around the first linear crossing $t=t_i$, we use the so-called $N$ matrix which relates the upper/lower band probability amplitudes at time right before the crossing $t=t_i^-$ to the upper/lower bands probability amplitudes at time right after the crossing $t=t_i^+$ \cite{SAN2010}. The $N$ matrix is a recasting of the exact solution of the time-dependent problem of a single linear avoided crossing \cite{LZ1932} in terms a scattering matrix, including the crucial phase information related to non-adiabatic processes. The $N$ matrix for the second linear crossing at $t=t_f$ is similarly defined, and they are given by
\beq
N_{t=t_i}=\left(\begin{array}{cc}\sqrt{1-P_{LZ}}e^{-i\varphi_S}&-\sqrt{P_{LZ}}\\ \sqrt{P_{LZ}}&\sqrt{1-P_{LZ}}e^{i\varphi_S} \end{array} \right)
\eeq
and
\beq
N_{t=t_f}=\left(\begin{array}{cc}\sqrt{1-P_{LZ}}e^{-i\varphi_S}&\sqrt{P_{LZ}}\\ -\sqrt{P_{LZ}}&\sqrt{1-P_{LZ}}e^{i\varphi_S} \end{array} \right)=(N_{t=t_i})^T,
\eeq
where $\varphi_S=\pi/4+\delta (\ln \delta -1)+\textrm{arg} \Gamma(1-i \delta)$ is the phase acquired upon being reflected at a LZ crossing (the so-called Stokes phase). The transpose relation between the two $N$ matrices is related to the fact that the notion of upper and lower bands are inverted for the first and second avoided crossings in $H_{\chi,\mu,\tau}(t)$. To give an example on how to read the $N$ matrix, according to $N_{t=t_i}$, an initial state right before the crossing $|\psi(t_i^- )\rangle=a_- |\psi_- (t_i^-)\rangle    +a_+  |\psi_+ (t_i^-)\rangle$ will be transformed to a final state right after the crossing as
\beqn
|\psi(t_i^+ )\rangle&=&\bigl(  \sqrt{1-P_{LZ}}e^{i\varphi_S}\,a_- +   \sqrt{P_{LZ}}\, a_+\bigr) |\psi_- (t_i^+)\rangle\nn\\
&& \!\!\!\!\!\!\!\!\!\!\!\!\!\!+\bigl(-\sqrt{P_{LZ}}\, a_- + \sqrt{1-P_{LZ}}e^{-i\varphi_S}\,a_+\bigr) |\psi_+ (t_i^+)\rangle.
\eeqn
Note that these amplitudes here do not contain any adiabatically accumulated phase. In this work we adopt the viewpoint that everything non-adiabatic is described by $N$ matrices, while everything adiabatic will be in phases acquired in between tunneling events \cite{footnote1}. Drawing the analogy with optical Mach-Zehnder interferometer, such an $N$ matrix characterizes a linearly avoided crossing as a beam splitter of transmission $P_{LZ}$ and reflection phase $\varphi_S$ (see, e.g., \cite{Zeilinger1981,Holbrow2002}).

In traversing two linear crossings in succession (the St\"uckelberg interferometer), we take the product of the two matrices $N_{t=t_f}\,N_{t=t_i}$ and read off the final non-adiabatic phase accumulated for the two possible paths. However, the amplitude for each path also contains a phase accumulated during the adiabatic evolution. Therefore, the amplitude $\textrm{Amp}_+$ for the upper path is:
$$
\textrm{Amp}_+=-\sqrt{P_{LZ}}\times e^{i\varphi_+}\times \sqrt{1-P_{LZ}}e^{-i\varphi_S}
$$
It is the product of three terms (see, for example, Ref. \onlinecite{FLM2012}). The first $-\sqrt{P_{LZ}}$ is the probability amplitude to tunnel from the lower to the upper band at the first crossing, and the third $\sqrt{1-P_{LZ}}e^{-i\varphi_S}$ is the probability amplitude not to tunnel (i.e. to stay in the upper band) at the second crossing. The second term $e^{i\varphi_+}$ is the complex exponential of the total phase of the adiabatic motion between the two crossings at $t_i$ and $t_f$ given by
\beqn
\varphi_+ &=&-\int_{t_i}^{t_f}dt E_+(t)+\int_{t_i}^{t_f}dt \langle \psi_+|i\partial_t|\psi_+\rangle \nn\\
&&+ \textrm{arg}\langle \psi_+(t_i)|\psi_+(t_f)\rangle\nn\\
&=&-\int_{t_i}^{t_f}dt E_+(t)+\Gamma_+.
\label{varphip}
\eeqn
The total adiabatic phase is itself the sum of three terms: a dynamical phase, a line integral of a Berry connection along an open-path and a projection (or geodesic) closure (i.e., the argument of an overlap between two eigenstates). While the dynamical phase depends on the bandstructure, the latter two depends on the band eigenstates along the path. Note that the path is open in the parameter space and that, in addition, the initial and final states are not proportional to each other.

The reason for the projection closure contribution can be understood in the adiabatic theory for the upper path between $t_i$ and $t_f$. Take an initial condition $|\psi(t_i)\rangle = |\psi_+(t_i)\rangle$ and compute $|\psi(t>t_i)\rangle$ using the adiabatic theory. This gives at the second crossing:
\beq
|\psi(t_f)\rangle = |\psi_+(t_f)\rangle e^{i\int_{t_i}^{t_f} dt [-E_+(t)+\langle \psi_+|i\partial_t|\psi_+\rangle]}
\label{finalstate}
\ee
Therefore, the adiabatically accumulated phase along the upper path starting at $t_i$ with $|\psi_+(t_i)\rangle$ and ending at $t_f$ with $|\psi_+(t_f)\rangle$ is the argument of
$$
\langle\psi_+(t_i)|\psi(t_f)\rangle = \langle\psi_+(t_i)|\psi_+(t_f)\rangle e^{i\int_{t_i}^{t_f} dt [-E_+(t)+\langle \psi_+|i\partial_t|\psi_+\rangle]}
$$
which is indeed (\ref{varphip}). Note that the open-path geometric phase $\Gamma_+$ is gauge invariant, thanks to the projection closure term \cite{SB1988,PS1998}. The expression for the geometric phase is well defined when the initial and final states are not orthogonal. We will come back to this point in section \ref{sec:massless} when the two states are orthogonal, and in sect. \ref{sec:geointer} we discuss its geometrical meaning.

The amplitude $\textrm{Amp}_-$ for the lower path is similarly given by the product of three terms (amplitude not to tunnel at the first crossing; adiabatically acquired phase; and amplitude to tunnel at the second crossing):
$$
\textrm{Amp}_-= \sqrt{1-P_{LZ}}e^{i\varphi_S}\times e^{i\varphi_-}\times \sqrt{P_{LZ}}
$$
where
\beqn
\varphi_-=-\int_{t_i}^{t_f}dt E_-(t)+\Gamma_-
\eeqn
is the total adiabatic phase accumulated by the particle traveling in the lower band from one crossing to the other.

The final transition probability is therefore
\beqn
P_f&=&|\textrm{Amp}_+ + \textrm{Amp}_-|^2\nn\\&=&4P_{LZ}(1-P_{LZ})\sin^2(\varphi_S + \frac{\varphi_- - \varphi_+}{2})\nn\\
&=&4P_{LZ}(1-P_{LZ})\sin^2(\varphi_S + \varphi_{dyn}/2+\varphi_g/2)
\eeqn
where the total phase (defined modulo $2\pi$) of the St\"uckelberg interferometer is the sum of a Stokes phase $\varphi_S$ (acquired during non-adiabatic tunneling events), a dynamical phase $\varphi_{dyn}=\int_{t_i}^{t_f}dt (E_+ - E_-)$ and a geometric phase $\varphi_g\equiv \Gamma_- - \Gamma_+$ given by
\beqn
\varphi_g&=& \int_{t_i}^{t_f}dt \langle \psi_-|i\partial_t|\psi_-\rangle + \textrm{arg}\langle \psi_-(t_i)|\psi_-(t_f)\rangle \nn\\
&-&\int_{t_i}^{t_f}dt \langle \psi_+|i\partial_t|\psi_+\rangle - \textrm{arg}\langle \psi_+(t_i)|\psi_+(t_f)\rangle,
\label{geomphas}
\eeqn
with the latter two are acquired during the adiabatic evolution in between the two crossings.

In short, we recover the expected \s interference structure in the transition probability with the overall double LZ tunnelings factor $2P_{LZ}(1-P_{LZ})$, a quantity determined solely by the adiabaticity parameter $\delta$. However, in the interference pattern, besides a phase modulation related to the bandstructure (i.e., the dynamical phase $\varphi_{dyn}$ and the Stokes phase $\varphi_S$), it generally contains a non-trivial geometric phase contribution $\varphi_g$, which requires the knowledge of the band eigenstates that is beyond the bandstructure. The appearance of this geometric phase is surprising at first sight, and it is sensitive to the pseudospin structure of the Hamiltonian $H_{\chi,\mu}(\vec{p})$ discussed in Sect.~\ref{sec:stuckhamil}.

\section{Dynamics of a quantum particle in the adiabatic basis}\label{sec:statement}
The derivation provided in the previous section, while physically appealing, requires a more careful justification. To introduce the solution methods in the following sections, we formulate the main time evolution equations of the complete $2\times 2$ Hamiltonian $H_{\chi,\mu,\tau}(t)$ in terms of the adiabatic basis. Let the state of the system be written as:
$$|\psi(t)\rangle=\sum_{\alpha=\pm} A_\alpha (t) e^{-i \int^t_0 dt' E_\alpha (t')}e^{i\int^t_0 dt' \langle \psi_\alpha |i\partial_t| \psi_\alpha\rangle}|\psi_\alpha(t)\rangle$$
with $E_-(t)=-E_+(t)$. Then the time-dependent Schr\"odinger equation gives:
\begin{eqnarray}
\dot{A}_+ &=& -\langle \psi_+|\dot{\psi}_-\rangle A_-  e^{i\int^t_0 dt' 2 E_+ }\nn\\
&&\times e^{i\int^t_0 dt' [\langle \psi_- |i\partial_t| \psi_-\rangle-\langle \psi_+ |i\partial_t| \psi_+\rangle]},\nn\\
\dot{A}_- &=& -\langle \psi_-|\dot{\psi}_+\rangle A_+ e^{-i\int^t_0 dt' 2 E_+}\nn\\
&&\times e^{-i\int^t_0 dt' [\langle \psi_- |i\partial_t| \psi_-\rangle-\langle \psi_+ |i\partial_t| \psi_+\rangle]},
\label{mainequations}
\end{eqnarray}
with the initial conditions $A_- (-\infty)=1$ and $A_+ (-\infty)=0$. We are interested in the final transition probability $P_f=|A_+(+\infty)|^2$. Band coupling occurs through $\langle \psi_+|\dot{\psi}_-\rangle$ which is the off-diagonal Berry connection $\mathcal{A}_{+,-}(t)\equiv \langle \psi_+| i\frac{d}{dt}|\psi_- \rangle$. For a derivation of the above equations using a scalar time-independent gauge, see appendix~\ref{sec:altder}.

In Ref. \cite{FLM2012}, we studied a similar set of time evolution equations with two avoided crossings, that actually corresponds to case \#1 (parallel trajectory) in Table II. Now both the off-diagonal Berry connection and the diagonal Berry connection $\langle \psi_\pm |i\partial_t| \psi_\pm \rangle $ generally permit a much richer analytic structure for the transition probability (see later in the adiabatic perturbation theory section). Specifically, case \#1 (and case \#6) is a special case where geometric corrections are absent due to vanishing of the diagonal Berry connection and the off-diagonal Berry connection being real. In general, they are non-zero and complex valued, and these will be shown to shift the the St\"uckelberg oscillations, i.e. to give rise to a geometric phase contribution to the final probability of the St\"uckelberg interferometer. Thus it generalizes our previous work Ref. \cite{FLM2012} in a crucial way.

\section{Dirac cones with same mass, opposite chirality and diagonal trajectory}\label{sec:twobands}
To proceed with explicit expressions for the time-dependent Hamiltonian, we take the specific case of two Dirac cones with the same mass, an opposite chirality and diagonal trajectory (case \#2 in Table~\ref{table3}), where we expect non-trivial geometrical effects. The rest of the paper will be devoted to this case, whereas in section \ref{sec:map} we give a summary of the results for the other cases.

We first study numerically the exact time evolution Eq.~(\ref{mainequations}) and compare with the result of the \s theory (see section \ref{sec:stucktheo}).
We then study the invariance of the geometric phase with respect to several choices. Only in the next section we use the adiabatic perturbation theory to derive the results analytically.

\subsection{\s regime}
The $2\times 2$ time-dependent Hamiltonian of case \#2 reads (in units such that $F_x=\hbar=2m=1$) \cite{LFM2014}
\beqn\label{eq:diaham}
H(t)=(t^2-\Delta_*)\sigma_x + c_y F_y t \sigma_y + M\sigma_z
\eeqn
with $E_\pm (t)=\pm[(t^2-\Delta_*)^2+c_y^2 F_y^2 t^2 + M^2]^{1/2}$. The band crossings occur at complex times $t$ such that $E_+(t)=0$. The \s regime corresponds to the limit in which the two tunneling events are well separated. In this limit, the two avoided linear crossings are at time $t_i\approx -\sqrt{\Delta_*}$ and $t_f=-t_i$ and are characterized by an energy gap of magnitude $2 [\Delta_* c_y^2 F_y^2+M^2]^{1/2}$, see Fig.~\ref{traj}b. The precise definition of the \s regime is that the time between tunneling events $\sim 2\sqrt{\Delta_*}$ should be much larger than their duration $\tau_{LZ}\sim \textrm{max}(\delta,\sqrt{\delta})/\sqrt{c_y^2F_y^2\Delta_*+M^2}$ with the adiabaticity parameter $\delta\sim (c_y^2F_y^2\Delta_*+M^2)/\sqrt{\Delta_*}$.  In practice, this means that $\Delta_* \gg \sqrt{c_y^2F_y^2\Delta_* + M^2}\geq c_y^2 F_y^2,M$, which we assume in the following.

\subsection{Numerics}
\begin{figure}[top]
\begin{center}
\includegraphics[width=8cm]{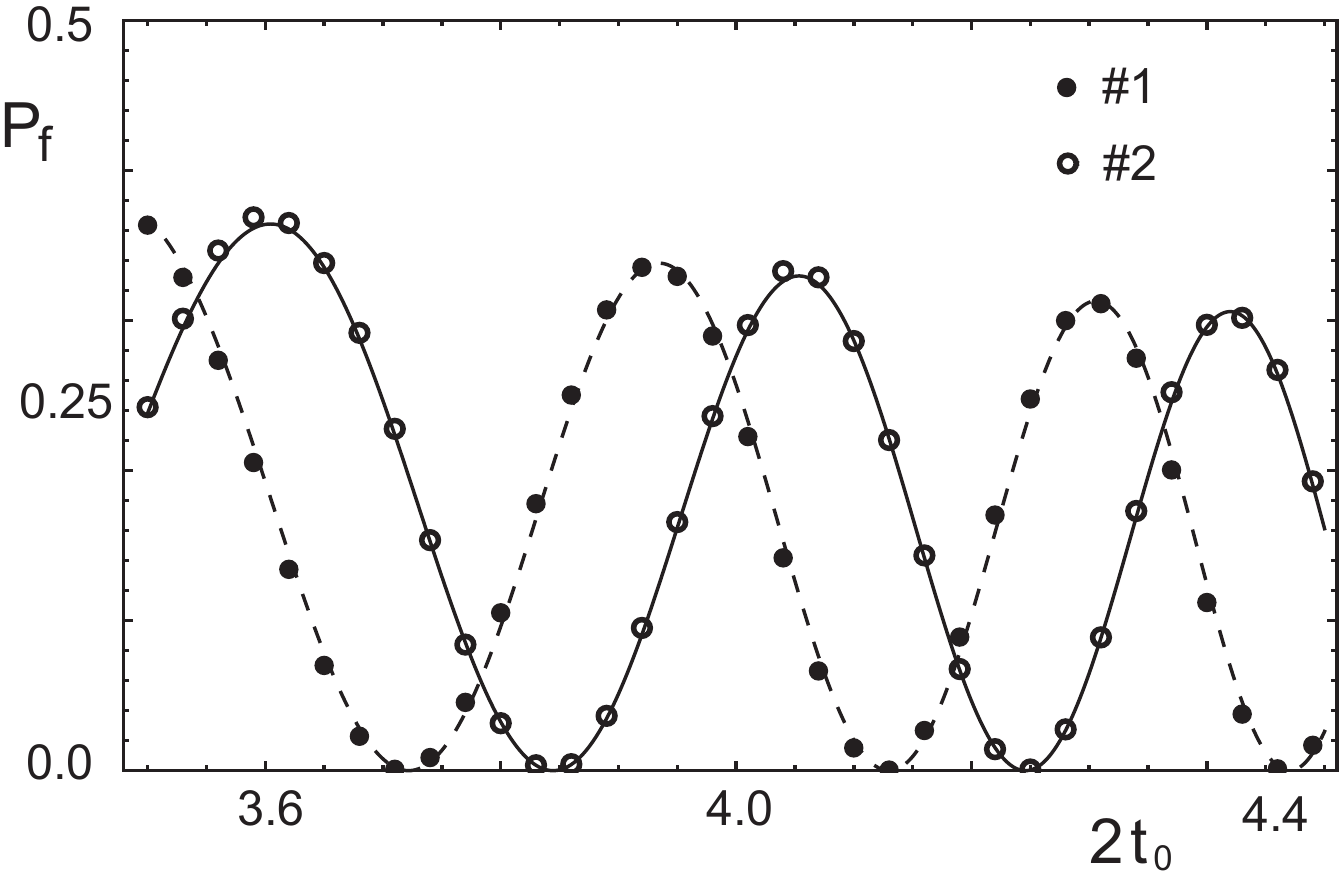}
\end{center}
\caption{Final transition probability $P_f$ as a function of the time interval $2 t_0 = 2\sqrt{\Delta_*}$ between the two Dirac cones for the time-dependent Hamiltonians of case \#1 and \#2 as indicated. The dots corresponds to the full numerical solution. The lines correspond to the St\"{u}ckelberg theory $P_f=4P_{LZ}(1-P_{LZ})\sin^2(\varphi_S+\varphi_{dyn}/2+\varphi_g/2)$. In case \#1 (dashed line) $\varphi_g=0$, whereas $\varphi_g=\Delta \varphi = -2\arctan \frac{c_y F_y
\sqrt{\Delta_*} }{M} \approx - \pi/4$ in case \#2 (full line), see subsect. \ref{sec:sec:geo}. The parameters are $c_y F_y = 0.14$, $M^2 = 0.35^2 - (c_yF_y t_0)^2$ and $c_yp_y=c_yF_yt_0$ such that the gap at the avoided crossings is $2\sqrt{M^2+(c_y F_y t_0)^2}=0.7$ in both cases.}
\label{fig:numfinitemass}
\end{figure}

The time evolution of the system is governed by Eq.~(\ref{mainequations}) using the Hamiltonian (\ref{eq:diaham}). We solve these equations numerically and compare also with the numerical result for a parallel trajectory (case \#1), shown in Fig.~\ref{fig:numfinitemass} as open and filled circles, respectively. We see that the interference fringes in the latter case agree well with the prediction of the \s theory in the absence of a geometric phase shift, namely $P_f=4P_{LZ}(1-P_{LZ})\sin^2\left(\varphi_S+\varphi_{dyn}/2\right)$, shown as dashed curve \cite{FLM2012}. On the other hand, there is an obvious mismatch between the cases \#1 and \#2 (open and filled circles), despite the fact that the two adiabatic spectra are identical, see Fig.~\ref{traj}b. However, the phase shift between the two cases is well accounted for by using the result of the \s theory including a geometric phase shift $\varphi_g=\Delta \varphi$ (see next section) with $P_f=4P_{LZ}(1-P_{LZ})\sin^2\left(\varphi_S+\varphi_{dyn}/2+\varphi_g/2\right)$, shown as the solid curve. This confirms numerically that the non-trivial geometrical shift provides an additional ingredient in the understanding of the \s phenomenon.

\subsection{Geometric phase}\label{sec:sec:geo}
We now examine more closely the geometric phase $\varphi_g$ of Eq. (\ref{geomphas}) for case \#2. Its explicit computation appears different for different choices  such as Hamiltonian bases or gauge choices for the associated adiabatic eigenstates. However, as we now show, the result is unique and well defined. 

\subsubsection{``Basis'' and gauge choices}
\begin{figure}
\begin{center}
\includegraphics[width=3.5cm]{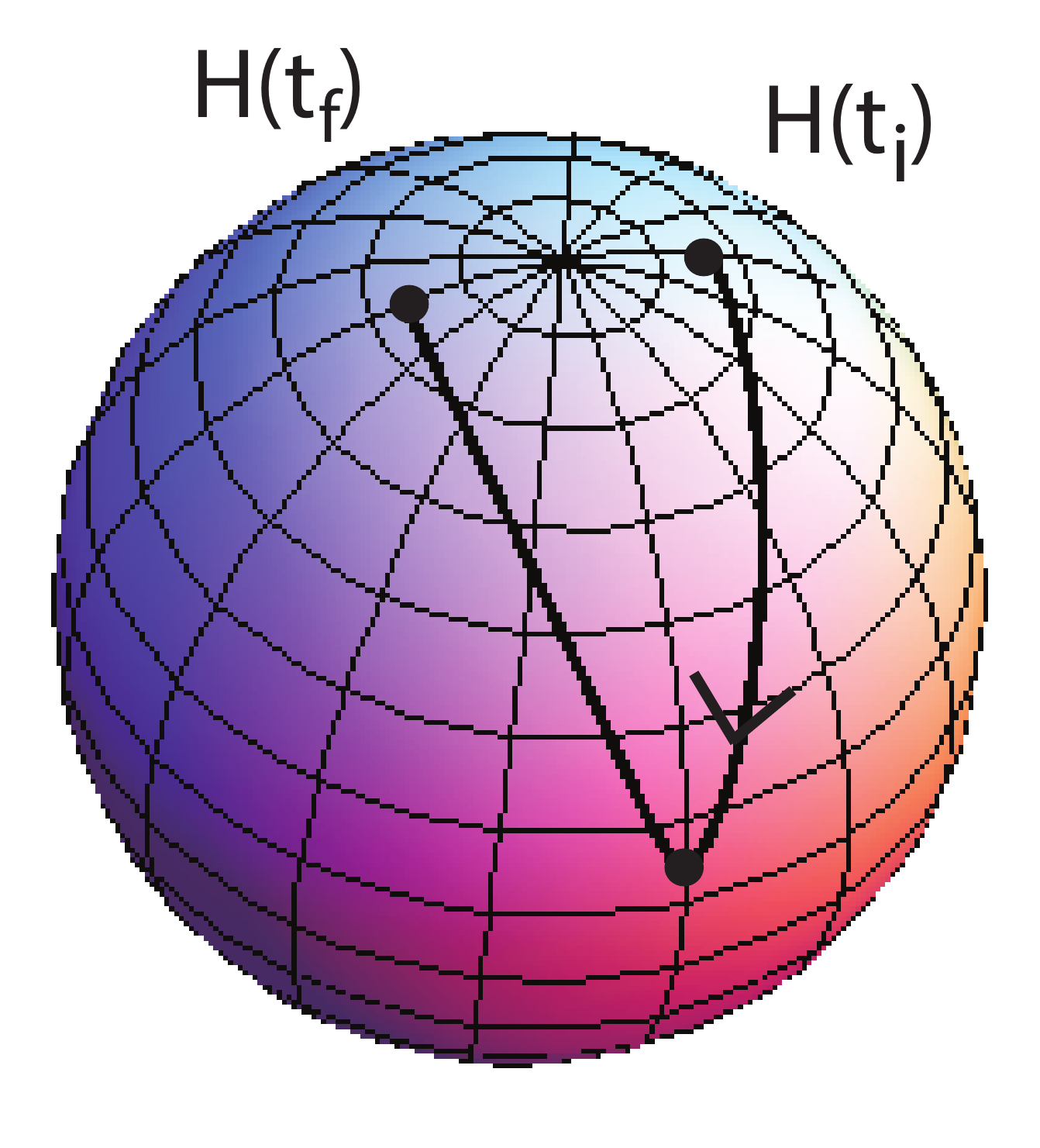}
\end{center}
\caption{Time evolution of the Hamiltonian $H(t)=E_+(t)\vec{n}(t)\cdot \vec{\sigma}$ represented by the curve traced by $\vec{n}(t)$ on the Bloch sphere from $t=t_i$ to $t=t_f$. The dot in the middle of the trajectory is at $t=0$.}\label{fig:bloch}
\end{figure}

For the case \#2 that we study, the Hamiltonian (\ref{eq:diaham}) given by
\beqn\label{parallelhamil}
H_{\tr{gr}}(t)=(t^2-\Delta_*)\sigma_x + c_y F_y t \sigma_y + M\sigma_z
\eeqn
is written in the ``natural" Pauli matrix basis when considering the tight-binding model for graphene consisting of two inequivalent Dirac points $D,D'$, (we call it the ``graphene (gr) basis'', see Eq. (1) in Ref.~\cite{LFM2014}). 
To help visualizing the time evolution of the Hamiltonian curve, we plot its trajectory on the Bloch sphere for the time in between $t_i$ and $t_f$, see Fig.~\ref{fig:bloch}(b). 

By performing a time-independent unitary rotation in pseudo-spin space the Hamiltonian can also be written as
\beqn\label{parallelhamillz}
H_{\tr{lz}}(t)= c_y F_y t \sigma_x + M \sigma_y + (t^2-\Delta_*)\sigma_z
\eeqn
We call it the ``Landau-Zener (lz) basis'', see Eq. (2) in Ref.~\cite{LFM2014}. It is just another representation for the Pauli matrices, with the property that the main time-evolution ($t^2-\Delta_*$) is on the matrix diagonal.
The Hamiltonian curve in this basis can be obtained from a global rotation of the Hamiltonian curve plotted for the graphene basis. It is convenient at this point to introduce the following angle
\beq
\phi_{\tr{lz}}(t_f)-\phi_{\tr{lz}}(t_i)= -2\arctan(c_y F_y \sqrt{\Delta_*}/M)\equiv \Delta \varphi
\label{deltavarphi}
\eeq
that will be shown to be equal to the geometric phase later. The subscript ``lz" reminds us that the parameters are obtained in the Landau-Zener basis.

\begin{table*}
\caption{Geometric phase $\varphi_g$ for Dirac cones with a constant mass, opposite chirality and a diagonal trajectory (case \# 2). It is computed as the sum of a Berry connection line integral $\Theta$ and a projection closure term $\Pi$ using different Hamiltonian bases and gauge choices for the eigenstates. Here $\Delta \varphi=-2\arctan(c_y F_y \sqrt{\Delta_*}/M)$. We see that $\varphi_g=\Delta \varphi$ in all cases.}\label{tablegaugebasis}
\vspace{0.25cm}
\begin{tabular}{c|c|c|c}
Basis/gauge & Berry connection line integral $\Theta$ & Projection closure $\Pi$ & Sum $\varphi_g$  \\
\hline
graphene/south (article \cite{LFM2014} choice) & $\Delta \varphi$ & $0$ & $\Delta \varphi$\\
graphene/north  & $\Delta \varphi$ & $0$ & $\Delta \varphi$\\
LZ/south & $2\Delta \varphi$ & $-\Delta \varphi$ & $\Delta \varphi$\\
LZ/north (parallel transport) & $0$ & $\Delta \varphi$ & $\Delta \varphi$\\
\end{tabular}
\end{table*}
Apart from these ``bases'' for the Hamiltonian $H(t)$, it is also necessary to specify the adiabatic eigenstates $|\psi_\pm(t)\rangle$ with a gauge choice. We consider two such choices called the south (S) pole gauge (when the multivaluedness of the eigenstates is at the south pole of the Bloch sphere) and the north (N) pole gauge, that combine to cover the whole parameter space of the Bloch sphere. The lower and upper band eigenstates in the two gauge choices are, respectively, given by
$$
\textrm{S:\ } |\psi_- \rangle=\left(\begin{array}{c}-e^{-i\phi} \sin\frac{\theta}{2}\\ \cos\frac{\theta}{2} \end{array}\right),\,\,|\psi_+ \rangle=\left(\begin{array}{c} \cos\frac{\theta}{2}\\ e^{i\phi}\sin\frac{\theta}{2} \end{array}\right),
$$
and
$$
\textrm{N:\ }|\psi_- \rangle=\left(\begin{array}{c}- \sin\frac{\theta}{2}\\ e^{i\phi} \cos\frac{\theta}{2} \end{array}\right),\,\,|\psi_+ \rangle=\left(\begin{array}{c} e^{-i\phi}\cos\frac{\theta}{2}\\ \sin\frac{\theta}{2} \end{array}\right).
$$
The two sets of eigenstates are related by a gauge transformation: $|\psi_{\pm}(t) \rangle \rightarrow  e^{\pm i\phi(t)}|\psi_{\pm}(t) \rangle$.

Independently of the gauge choice, one has $\langle \psi_\pm(t)|\vec{\sigma}|\psi_\pm(t)\rangle =\pm \vec{n}(t)$ which belongs to the unit sphere. Therefore the Bloch sphere can be seen as either representing the direction $\vec{n}(t)$ of the magnetic field specifying the Hamiltonian $H(t)=E_+(t)\vec{n}(t)\cdot \vec{\sigma}$ or as being the projective Hilbert space for the upper band (each point $\vec{n}=\langle \psi_+|\vec{\sigma}|\psi_+\rangle$ of the sphere represents a ray $e^{i\alpha} |\psi_+\rangle $ where $\alpha$ is an arbitrary phase).

\subsubsection{Computation of the geometric phase}
The geometric phase is the sum of two terms $\varphi_g=\Theta+\Pi$: (i) The line integral of the Berry connection $\Theta\equiv \int_{t_i}^{t_f}dt [\langle \psi_-|i\partial_t|\psi_-\rangle-\langle \psi_+|i\partial_t|\psi_+\rangle]$, which is $\int_{t_i}^{t_f}dt \dot{\phi}(1-\cos\theta)$ in the south pole gauge and $\int_{t_i}^{t_f}dt \dot{\phi}(-1-\cos\theta)$ in the north pole gauge. (ii) The projection closure $\Pi\equiv\textrm{arg} \langle \psi_- (t_i)|\psi_- (t_f)\rangle - \textrm{arg} \langle \psi_+ (t_i)|\psi_+ (t_f)\rangle$, which is $2 \textrm{arg}\left(e^{-i(\phi(t_f)-\phi(t_i))}\sin\frac{\theta(t_i)}{2}\sin\frac{\theta(t_f)}{2}+\cos\frac{\theta(t_i)}{2}\cos\frac{\theta(t_f)}{2}\right)$ in the south pole gauge and $2 \textrm{arg}\left(\sin\frac{\theta(t_i)}{2}\sin\frac{\theta(t_f)}{2}+\cos\frac{\theta(t_i)}{2}\cos\frac{\theta(t_f)}{2}e^{i(\phi(t_f)-\phi(t_i))}\right)$ in the north pole gauge.

Using Table~\ref{tableappendix} in appendix~\ref{sec:appenzeeman}, we compute separately the contributions $\Theta$ and $\Pi$ in the two gauge choices with the two Hamiltonian bases. The results are summarized in Table~\ref{tablegaugebasis}. We always take the limit of well separated tunneling events ($\Delta_* \gg c_y^2 F_y^2, M$), i.e. deep in the St\"uckelberg regime, at the end of the computation. The line integral of the Berry connection in the LZ basis simplifies in the St\"uckelberg regime in noting that the function
$\cos \theta_{\tr{lz}}\approx -1$ for all $t$ except very close to $t= t_{i,f}$ where $\cos \theta_{\tr{lz}} =0$. So we can approximate the integral $\int_{t_i}^{t_f} dt \dot{\phi_{\tr{lz}}} \cos \theta_{\tr{lz}} \approx - \int_{t_i}^{t_f} dt \dot{\phi_{\tr{lz}}}= -\Delta \varphi$.

From the results of Table~\ref{tablegaugebasis}, we see that the terms $\Theta$ and $\Pi$, obtained after explicit computations of different integrals and expressions, each can assume different values from one basis/gauge choice to another but the sum of the two is an invariant (modular $2\pi$) given by $\varphi_g=\Delta \varphi$. We thus show explicitly that the geometric phase $\varphi_g=\Gamma_--\Gamma_+$ given by the difference between two open-path geometric phases $\Gamma_\pm$ is a well defined quantity: it does not depend on a choice of basis or on a gauge choice.

\section{Adiabatic perturbation theory: proof of the geometric phase}\label{sec:adiatheo}
Here we use adiabatic perturbation theory (APT) to analyze the \s phenomenon. In order to solve Eq. (\ref{mainequations}), we follow the calculation done in \cite{FLM2012} for the case of a parallel trajectory with $M_z(\vec{p})=0$ (case \#1) and adapt it to the problem of a diagonal trajectory with $M_z(\vec{p})=M$ (case \#2), see Hamiltonian (\ref{eq:diaham}). The goal is to use APT to compute the tunneling probability directly for the \s interferometer and to prove that, in the adiabatic limit in which $\delta\to \infty$, $\varphi_S\to 0$ and $P_{LZ}=e^{-2\pi \delta}\to 0$, the total tunneling probability is $P_f\approx 4 P_{LZ}\sin^2[\frac{\varphi_{dyn}+\varphi_g}{2}]$ with a geometric phase given by $\varphi_g=\Gamma_--\Gamma_+$.

\subsection{First order adiabatic perturbation theory}
We start from equations (\ref{mainequations}). First order perturbation theory means $|A_- (t)|\approx 1\gg |A_+ (t)|$ and therefore $\dot{A}_- \approx 0$ so that $A_- (t)\approx 1$ at all $t$. As a consequence, we are left with one equation to solve in this approximation,
\beqn
A_+(+\infty)&=& -\int_{-\infty}^{\infty} dt \langle \psi_+|\dot{\psi}_-\rangle   e^{i\int^t_0 dt' 2 E_+} \nn\\
&&\times e^{i\int^t_0 dt' [\langle \psi_- |i\partial_t| \psi_-\rangle-\langle \psi_+ |i\partial_t| \psi_+\rangle]},\label{amp1}
\eeqn
in order to compute the transition probability $P_f=|A_+(+\infty)|^2$.

We first note that by performing a gauge transformation, $|\psi_\alpha (t)   \rangle \rightarrow e^{i \zeta_\alpha (t)}  |\psi_\alpha (t)   \rangle$, the amplitude transforms according to $A_+(+\infty)\rightarrow e^{i (\zeta_-(0)-\zeta_+(0)) } A_+(+\infty)$. Thus it differs by a constant phase, which has no consequence in the tunneling probability. Without loss of generality we use the eigenstates in the south pole gauge throughout this section. This leads to a band coupling expression $-\langle \psi_+|\dot{\psi}_-\rangle = (1/2)(\dot{\theta}-i\dot{\phi}\sin \theta)e^{-i\phi}$ and also the Berry connection $\langle \psi_- |i\partial_t| \psi_-\rangle-\langle \psi_+ |i\partial_t| \psi_+\rangle=\dot{\phi}(1-\cos\theta)$. We then obtain for the amplitude
\beqn\label{amp11}
A_+(+\infty)&=& \int_{-\infty}^{\infty} dt  \f{\dot{\theta}-i\dot{\phi}\sin \theta}{2}\,e^{i\beta(t)}
\eeqn
where the total phase is $\beta(t) \equiv -\phi(t)  +\int_0^{t} dt' \dot{\phi}(1-\cos\theta)+\int_0^{t} dt'2E_+(t')$. The explicit form of the integrand with case \#2 of the Hamiltonian (\ref{parallelhamillz}) in the Landau-Zener basis can be written down with the help of Table~\ref{tableappendix} in Appendix~\ref{sec:appenzeeman}. As we will see, each term in the phase has its corresponding physical meaning as we already encountered in section \ref{sec:stucktheo}: the $e^{-i \phi}$ term will give rise to the projection closure when evaluated close to the two poles at $t\sim t_{i,f}$; $e^{i\int_0^t dt' \dot{\phi}(1-\cos\theta)}$ the line integral of the Berry connection and $e^{i\int^t_0 dt' 2 E_+}$ will give the dynamical phase.

To evaluate the expression Eq.~(\ref{amp11}), we perform a contour integration in the complex time plane. As the computation of in the complex plane is quite lengthy, we give the details in the appendix \ref{sec:appintcomplex} and directly discuss the results. The amplitude $A_+(+\infty)$ is given as the sum of two dominant residues coming from poles at $t_1\approx \sqrt{\Delta_*}+ic_y F_y/2$ and $t_4=-t_1^*$. It reads
\beqn\label{res}
A_+(+\infty)=-\frac{\pi}{3}(e^{- \textrm{Im}\beta_1+i \textrm{Re}\beta_1}-e^{-\textrm{Im}\beta_4+i \textrm{Re}\beta_4})
\eeqn
where we defined $\beta_{1,4}\equiv\beta(t_{1,4})$.

To simplify further, we note that for the imaginary part of $\beta_{1,4}$, we have $\textrm{Im}\beta_1=\textrm{Re}\int_0^{\textrm{Im}t_1}dv (2 E_+- \dot{\phi}\cos \theta)|_{t'=\textrm{Re}t_1+iv}\approx \int_0^{\textrm{Im}t_1}dv (2 E_+- \dot{\phi}\cos \theta)|_{t'=\textrm{Re}t_1}$  when $\textrm{Re}t_1= \sqrt{\Delta_*}\gg \textrm{Im}t_1\approx c_yF_y/2$. Similarly for $\textrm{Im}\beta_4\approx \int_0^{\textrm{Im}t_1}dv (2 E_+- \dot{\phi}\cos \theta)|_{t=-\textrm{Re}t_1}$. But the terms $E_+$, $\dot{\phi}$ and $\cos \theta$ are even function of $t'$ and therefore $\textrm{Im}\beta_1=\textrm{Im}\beta_4$.
We then have
\beqn
A_+(+\infty)=-\frac{\pi}{3}e^{-\textrm{Im}\beta_1}e^{i\frac{\textrm{Re}\beta_1 + \textrm{Re}\beta_4}{2}}2i\sin(\frac{\textrm{Re}\beta_1 - \textrm{Re}\beta_4}{2})
\eeqn
and
\beqn
P_f&=&|A_+(\infty)|^2\approx (\frac{\pi}{3})^2 4e^{-2\textrm{Im}\beta_1} \sin^2 (\frac{\textrm{Re}\beta_1 - \textrm{Re}\beta_4}{2})\nn\\
&\to& 4e^{-2\textrm{Im}\beta_1} \sin^2 (\frac{\textrm{Re}\beta_1 - \textrm{Re}\beta_4}{2}),
\label{finalresult}
\eeqn
where in the last line we use the standard procedure to resolve the ``$\pi/3$ problem'' \cite{D1960,Davis1976,Berry1982}. We now recognize the general St\"uckelberg probability structure $P_f=4P_{LZ}(1-P_{LZ})\sin^2[\varphi_S+(\varphi_{dyn}+\varphi_g)/2]$ with $P_{LZ}$ replaced by $e^{-2\textrm{Im}\beta_1}$ and the total phase $\varphi_S+(\varphi_{dyn}+\varphi_g)/2$ replaced by $\textrm{Re}\beta_1 - \textrm{Re}\beta_4$. In the adiabatic limit, we always have that $1-P_{LZ}\approx1$ and $\varphi_S\rightarrow 0$.

\subsection{St\"uckelberg formula with geometric correction}
We consider separately the real and imaginary parts of $\beta_{1,4}$. For the real part, we have $\textrm{Re}\beta_1= \textrm{Re}\alpha_1-\pi/2- \textrm{Re}\int_0^{t_1} dt \dot{\phi_{\tr{lz}}}\cos\theta_{\tr{lz}}\approx  \textrm{Re}\alpha_1-\pi/2+\int_0^{t_f} dt \dot{\phi_{\tr{lz}}}\approx  \textrm{Re}\alpha_1+\phi_{\tr{lz}}(t_f)-\pi$, and $\textrm{Re}\beta_4\approx \textrm{Re}\alpha_4+\phi_{\tr{lz}}(t_i)-\pi$. We finally get
\beqn\textrm{Re}\beta_1 - \textrm{Re}\beta_4&\approx& 2\int_{t_i}^{t_f}dt E_+(t)+\phi_{\tr{lz}}(t_f)-\phi_{\tr{lz}}(t_i)\nn\\&=&\varphi_{dyn}+\Delta \varphi,\eeqn
as expected for the phase of the St\"uckelberg interferometer in the adiabatic limit in which the Stokes phase vanishes.

We therefore find that indeed the extra phase $\varphi_g$ in the \s oscillations, which we previously identified to the two-band open-path geometric phase $\Gamma_- -\Gamma_+$, is also equal to $\Delta \varphi$, which is a phase difference accumulated at each tunneling event. Note that the two derivations are quite different: on the one hand, the extra phase appears as being accumulated during the adiabatic evolution along each band, and on the other hand, it seems to be captured during tunneling events.

For the imaginary part, we have $\textrm{Im}\beta_1=\textrm{Im}\alpha_1-\textrm{Im}\int_0^{t_1}dt \dot{\phi}\cos \theta$. First $\textrm{Im}\alpha_1\approx\frac{\pi}{4}c_y^2F_y^2\sqrt{\Delta_*}$. We recognize the adiabaticity parameter $\delta=\frac{c_y^2 F_y^2 \Delta_* + M^2}{4\sqrt{\Delta_*}}\approx \frac{c_y^2 F_y^2 \sqrt{\Delta_*}}{4}$ and the LZ probability $P_Z=e^{-2\pi\delta}\approx e^{-\pi c_y^2 F_y^2 \sqrt{\Delta_*}/2}\approx e^{-2\textrm{Im}\alpha_1}$. This is not surprising as $e^{-2\textrm{Im}\alpha_1}=e^{-2}\textrm{Im}\int_0^{t_1}dt [E_+(t)-E_-(t)]$ is the general expression of Dykhne for the tunneling probability in the adiabatic limit \cite{D1960}. Second, there is a small correction to the LZ probability coming from the imaginary part of the ``geometric phase'' $-\textrm{Im}\int_0^{t_1} dt \dot{\phi_{\tr{lz}}}\cos \theta_{\tr{lz}}$. This is similar to the geometrical correction for the tunneling probability found for a single avoided crossing by Berry in 1990 \cite{B1990}. This is a small correction that we neglect in the following.

Eventually, in the spirit of the Dykhne-Davis-Pechukas formula \cite{D1960, Davis1976}, we can propose a heuristic generalization of the result (that we obtained in the adiabatic limit), which should work well also in the diabatic and in the intermediate force regimes:
\begin{equation}
P_f\approx 4e^{-2\textrm{Im}\beta_1}(1-e^{-2\textrm{Im}\beta_1}) \sin^2 (\frac{\textrm{Re}\beta_1 - \textrm{Re}\beta_4}{2}+\varphi_{S})
\end{equation}
This was called the ``modified St\"uckelberg formula'' in \cite{FLM2012} with the extra extension that it now also includes geometrical effects. Note that, here, $\frac{\textrm{Re}\beta_1 - \textrm{Re}\beta_4}{2}\neq \textrm{Re}\beta_1$ due to the geometric phase, in contrast to the case found in \cite{FLM2012}.

\subsection{Hamiltonian in the graphene basis}\label{hamgrbasis}
Finally, we also want to verify that the result is independent of the Hamiltonian basis used. Indeed, the same analysis can be repeated for the same trajectory and parameters with the Hamiltonian (\ref{parallelhamil}) in the graphene basis. Following the procedures as in appendix \ref{sec:appintcomplex}, we arrive at the final probability amplitude
\beqn
|A_+|^2&=& 4e^{-2\textrm{Im}\beta_{\tr{gr},1}} \cos^2\biggl( \frac{\textrm{Re}\beta_{\tr{gr},1}-\textrm{Re}\beta_{\tr{gr},4}}{2}\biggr),\nn\\
\eeqn
noting the $\cos^2(\ldots)$ dependence in this basis (rather than $\sin^2(\ldots)$ dependence) due to the difference in the sign of the residue contributions. The subscript ``gr" reminds us that the expressions are obtained in the graphene basis. In particular, its argument is given by
\beqn
\textrm{Re}\beta_{\tr{gr},1}-\textrm{Re}\beta_{\tr{gr},4}&=&\varphi_{dyn}-\int_{t_i}^{t_f}dt' \dot{\phi_{\tr{gr}}}\cos\theta_{\tr{gr}}.
\eeqn
The last integral can be evaluated to give $-\int_{t_i}^{t_f}dt' \dot{\phi_{\tr{gr}}}\cos\theta_{\tr{gr}}\approx\Delta \varphi +\pi$ (see appendix \ref{sec:appenx}). And so the extra ``$\pi$-shift" in the second term brings us back to the same result as Eq.~(\ref{finalresult}). In other words, we arrive at the same gauge invariant geometric contribution $\Delta \varphi$ in the St\"{u}ckelberg interferometer, in the adiabatic limit. 

Looking back at Table~\ref{tablegaugebasis}, we see that the two bases result in Berry connection and projection closure terms that are quite different. It is \textit{a priori} a result based on a heuristic derivation of the St\"{u}ckelberg theory. Here with the adiabatic perturbation theory in the two bases, we prove that the final gauge invariant observable is indeed $\varphi_g=\Delta \varphi$.

\section{Massless Dirac cones with opposite chirality and diagonal trajectory}\label{sec:massless}
The massless limit of case \#2 (Dirac cones with \textit{zero} mass) deserves special attention. By restricting to $M=0$, the Hamiltonian curve is restricted to evolve on a great circle of the Bloch sphere (i.e. on the equator in the graphene basis as $\theta_{\tr{gr}}(t)=\pi/2$) with only two out of three Pauli matrices appearing in the Hamiltonian $H(t)$, despite the fact that the spectrum remains gapped $E_{\pm}(t)=\pm[(t^2-\Delta_*)^2 +c_y^2 F_y^2 t^2]^{1/2}$. This is a limit which is often referred to as possessing a chiral or sublattice symmetry in the graphene literature (in the graphene basis, $\sigma_z$ plays the role of a chiral operator as it squares to 1 and anticommutes with the Hamiltonian).

It follows that the open-path geometric phase $\Gamma_-$ is either equal to $0$ or to $\pi$ or is ill-defined (the last case being when the initial and end points are antipodal). Because $\Gamma_+=-\Gamma_-$, we find that $\Gamma_- -\Gamma_+=2\Gamma_-=0$ modulo $2\pi$ (except in the antipodal case). Therefore, it would seem that $\varphi_g=0$ in the massless case. Actually, the initial and end points of the Hamiltonian curve for case \#2 with $M\to 0$ move to $\theta_{\tr{gr}}(t_i)=\theta_{\tr{gr}}(t_f)\to \pi/2$ and $\phi_{\tr{gr}}(t_{i})=0, \phi_{\tr{gr}}(t_{f})=\pi$. On the Bloch sphere, thus, they are positioned precisely at antipodal position, which results in this case in an ill-defined expression for the geometric phase (see, however, Ref. \cite{Wong2005}). We therefore devote this section to the study of this special limit using several techniques.

Following the solution methods in the last two sections, we first numerically solve the time evolution equation. Then we use the adiabatic perturbation theory in two ways: first, we take the massless limit $M\to 0$ of the ``massive" case of Eq. (\ref{finalresult}). Second, we redo the adiabatic perturbation theory working directly with $M=0$.

\subsection{Numerics}
\begin{figure}[top]
\begin{center}
\includegraphics[width=7cm]{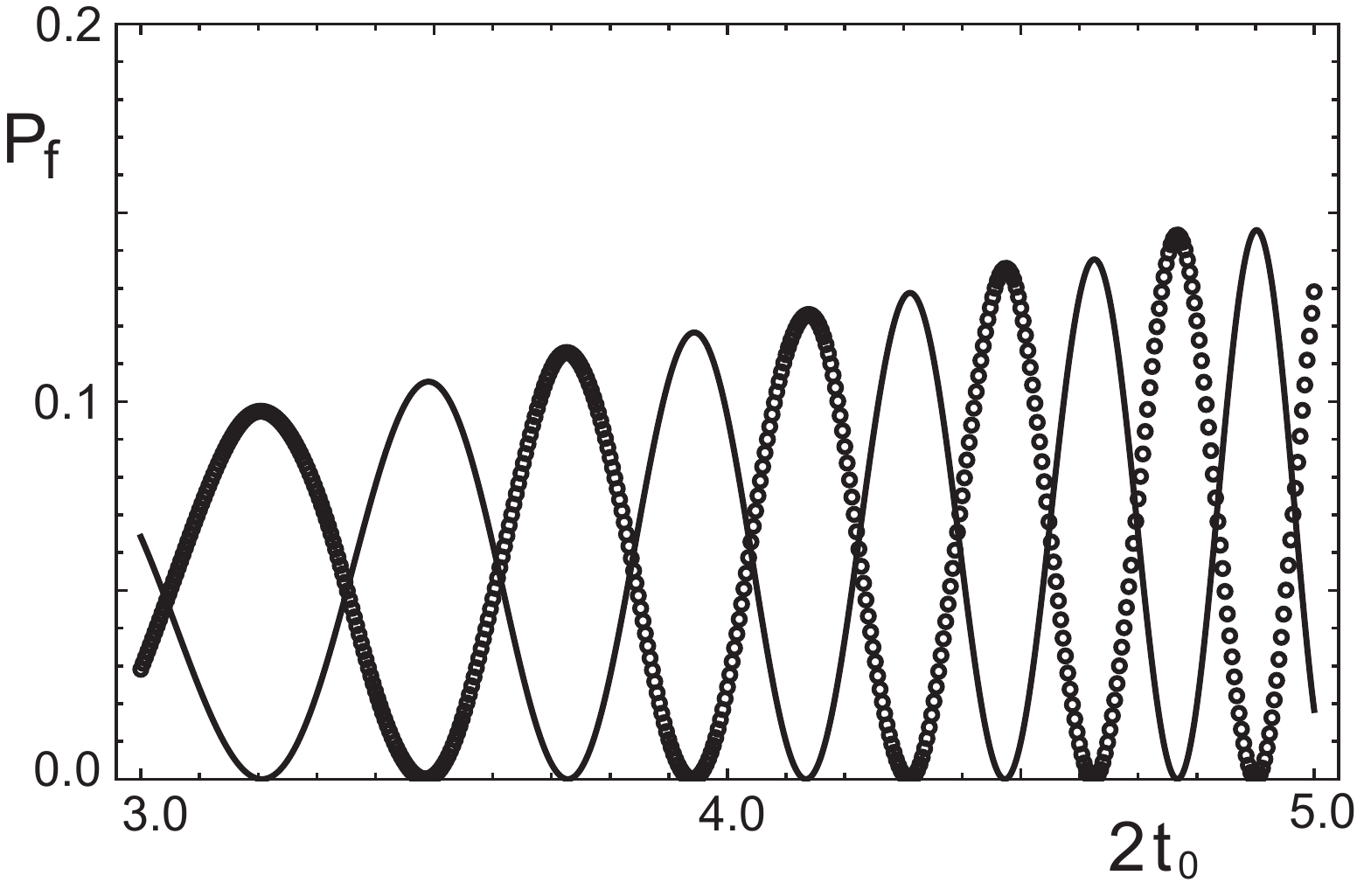}
\end{center}
\caption{Final transition probability $P_f$ as a function of the time interval $2 t_0 = 2\sqrt{\Delta_*}$ between the two Dirac cones for the case of \#2 with $M=0$. The open circles are the full numerical solution. The solid curve corresponds to \s theory without the geometric phase $P_f=4P_{LZ}(1-P_{LZ})\sin^2(\varphi_S+\varphi_{dyn}/2)$. The parameters are $c_y F_y=0.1$ with the gap at the avoided crossings given by $2c_y F_y t_0$. There is a clear $\pi$ shift.}
\label{fig:numerics}
\end{figure}

We numerically solve the time-dependent Schr\"odinger equation (\ref{mainequations}) for case \#2 with $M=0$ in the \s regime, see Fig.~\ref{fig:numerics} with open circles. As a reference, we compare it with the \s theory in the absence of a geometric phase shift, namely $P_f=4P_{LZ}(1-P_{LZ})\sin^2\left(\varphi_S+\varphi_{dyn}/2\right)$, shown as the solid curve. We recognize there is a clear $\pi$-shift difference in the phases in between the two.

\subsection{Adiabatic perturbation theory in the $M\to 0$ limit}
The APT result for case \#2 with a finite mass is $P_f\approx 4e^{-2\textrm{Im}\beta_1}\sin^2\left(\frac{\textrm{Re}\beta_1-\textrm{Re}\beta_4}{2}\right)$
with $\textrm{Re}\beta_1-\textrm{Re}\beta_4=\varphi_{dyn}+\Delta \varphi$, $\Delta \varphi=-2\arctan(c_yF_y\sqrt{\Delta_*}/M)$ and $\textrm{Im}\beta_1=\textrm{Im}\alpha_1-M/(2\Delta_*)$. We also noted that in the Landau-Zener basis, when $M\to0$ the $\alpha_6$ pole does not contribute to $A_+$ either because of the vanishing $M^3$ pre-exponential factor in the residue, see the paragraph before Eq. (\ref{res}). Therefore we can still use the finite $M$ form of
$P_f$ and take $M\to 0$ there, obtaining $\Delta \varphi =-\pi$, $\textrm{Im}\beta_1=\textrm{Im}\alpha_1$ and eventually:
$$
P_f\approx 4e^{-2\textrm{Im}\alpha_1}\sin^2\left(\frac{\textrm{Re}\alpha_1-\textrm{Re}\alpha_4-\pi}{2}\right)
$$
Showing the presence of a $\pi$-shift in the interferences.

\subsection{Adiabatic perturbation theory directly with $M=0$}
Here it is easier to work in the graphene basis (because in the LZ basis, the path on the Bloch sphere passes trough the poles):
$$
H_{\tr{gr}}(t)=(t^2-\Delta_*)\sigma_x+c_yF_y t \sigma_y
$$
and we will use the south pole gauge eigenstates. Following the same technique of complex integration (see appendix \ref{sec:appintcomplex}), we obtain for the final amplitude
as given by the sum of the residues (see appendix \ref{sec:appintcomplex2})
\beqn
A_+(+\infty)&=&-2\pi i(\frac{1}{6}e^{i\alpha_1}+\frac{1}{6}e^{-i\alpha_1^*})\nn\\&=&-\frac{\pi}{3}ie^{-\textrm{Im}\alpha_1}2\cos(\textrm{Re}\alpha_1),
\eeqn
and therefore giving the transition probability
\beqn
P_f&=&|A_+(+\infty)|^2\nn\\&\to& 4e^{-2\textrm{Im}\alpha_1}\cos^2(\textrm{Re}\alpha_1-\textrm{Re}\alpha_1)\nn\\
&=&4e^{-2\textrm{Im}\alpha_1}\sin^2\left(\frac{\textrm{Re}\alpha_1-\textrm{Re}\alpha_4+\pi}{2}\right),
\eeqn
in agreement with the previous methods showing the presence of the extra $\pi$-shift.

In summary, we have shown that there can be a geometric phase also in the massless case (i.e. when the trajectory on the Bloch sphere is restricted to a great circle). We have found that this phase $\varphi_g$ is either $0$ (case \# 1, see \cite{FLM2012}) or $\pi$ (case \# 2).

\section{Geometric phase as a solid angle}\label{sec:geointer}
To complete our understanding of the phase shift $\varphi_g$, let us focus on the geometrical meaning of the expression
\beqn\label{openpath}
\Gamma=\int_{\mathcal{C}} dt \langle\psi(t) | i\partial_t|\psi(t)\rangle +\textrm{arg}[\langle \psi(t_i)|\psi(t_f)\rangle].
\eeqn
It is an open-path geometric phase because the initial and final states are not necessarily proportional. Samuel and Bhandari \cite{SB1988} supplemented the open-path line integral of the Berry connection (first term of Eq.~(\ref{openpath}) with a geodesic closure (second term) making the sum of the two gauge-invariant. The open-path geometric phase was measured, for example, in neutron interferometry \cite{Wagh1998}. On the Bloch sphere, the quantity $\Gamma$ is equal to half the solid angle (or area) subtended by the path $\mathcal{C}$ closed by the shortest geodesic connecting $|\psi(t_f)\rangle $ and $|\psi(t_i)\rangle$ \cite{PS1998}. This is known as the geodesic rule (see a simple proof in appendix \ref{sec:geodrule}). With this picture, we understand why the projection closure term becomes ill defined when the initial and final states sit at antipodal position (when they are orthogonal) - there is no unique geodesic line connecting the two points.

\begin{figure}[ht]
\begin{center}
\includegraphics[width=7.cm]{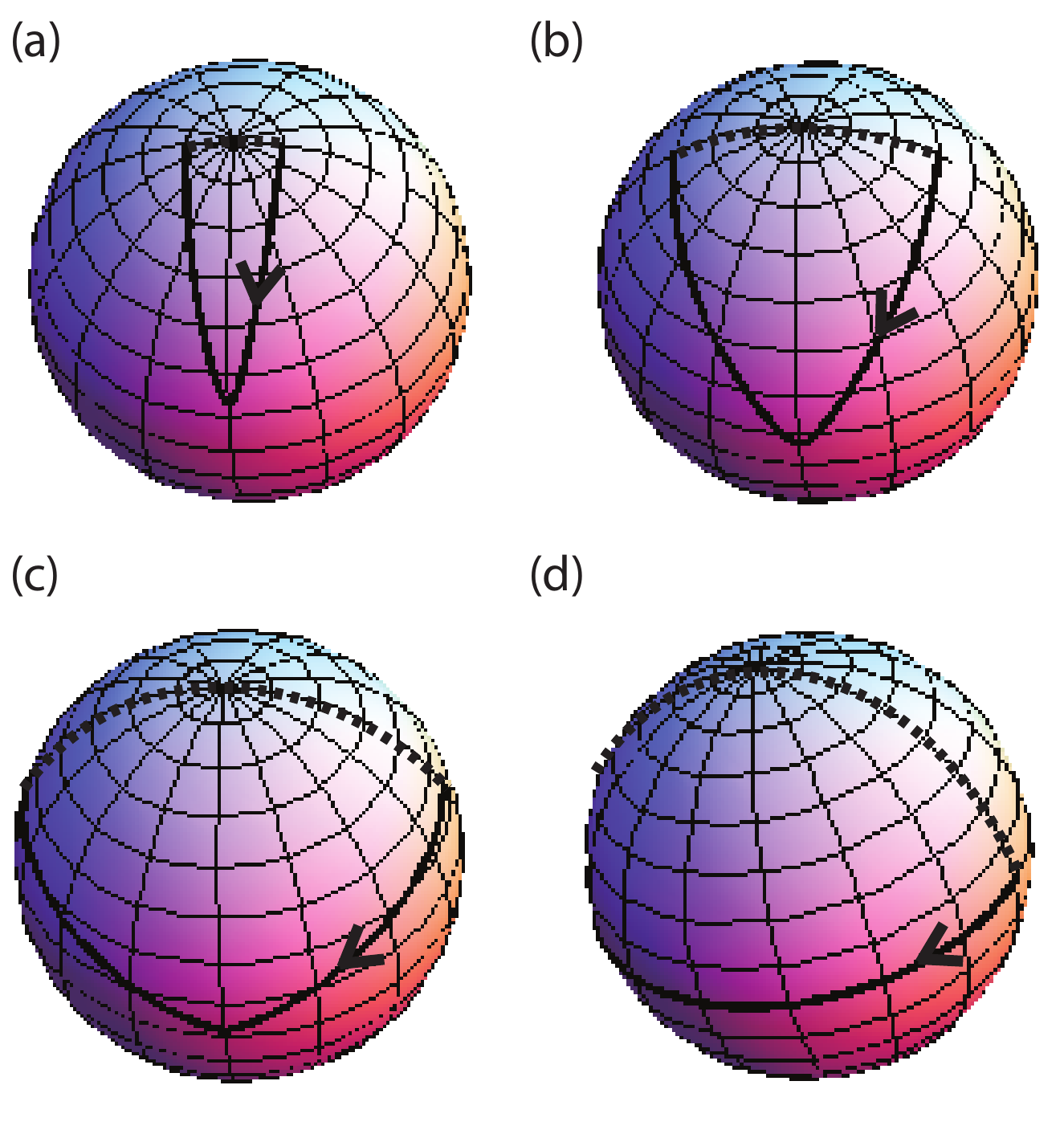}
\end{center}
\caption{Trajectories (full line) of the Hamiltonian (\ref{parallelhamil}) as the dimensionless mass parameter $M/(c_yF_y\sqrt{\Delta_*})$ decreases from 0.2 in (a) to $0$ in (d). The dotted line is the shortest geodesic line connecting positions of the initial and the final points of the Hamiltonian curve. We see that the area of the enclosed region evolves smoothly to the value $\pi$ for $M=0$.}\label{fig:geome}
\end{figure}

In the St\"{u}ckelberg interferometer the geometric phase shift is given by $\varphi_g=\Gamma_- -\Gamma_+$. Focusing on the lower band geometric phase $\Gamma_{-}$, typical trajectories on the Bloch sphere -- according to the Hamiltonian curve (\ref{parallelhamil}) in the graphene basis with a finite mass $M$ -- are shown in Fig. \ref{fig:geome}. The shortest geodesic path is indicated as the dotted line. The latter is part of the great circle passing the north pole, since the initial and final points lie on the opposite end of the azimuthal angle, i.e., $\phi=0, \pi$. According to the geometric phase formula, the enclosed area is then given by $|\varphi_g|=|\Gamma_--\Gamma_+|=2|\Gamma_-|$, which is the full solid angle, rather than half the solid angle due to two equal contributions from both the upper and lower bands (a model with particle-hole symmetry). With this interpretation, we can understand the massless limit $M\rightarrow 0$ quite naturally as the limiting area spanning a quarter of the Bloch sphere $|\varphi_g|=2 \arctan (c_y F_y \sqrt{\Delta_*}/M)\rightarrow \pi$. In Fig. \ref{fig:geome}, we show the evolution of the enclosed area as the mass parameter decreases.

\section{Phase shift for the eight double cone interferometers}\label{sec:map}
\begin{table*}
\caption{The phase shift $\Delta \varphi$ for the 8 different time-dependent Hamiltonians $H_{\chi,\mu,\tau}(t)=X_{\chi,\tau}(t)\,\sigma_x+Y_{\chi,\tau}\,(t)\sigma_y+Z_{\mu}(t)\,\sigma_z $ with $(\chi,\mu,\tau)=(\pm,\pm,\pm)$, where $\tau=+1$ for a parallel trajectory $\vec{p}(t)=(F_x t,p_y)$ or $\tau=-1$  for a diagonal trajectory $\vec{p}(t)=(F_x t,F_y t)$. Here we reintroduced $F_x$ and $m$ and do not take units such that $F_x=\hbar=2m=1$. We also give the closest ``distance'' to the Dirac points $\mathcal{D}\equiv |Y_{\chi,\tau}(t_0)|$ where $t_0\approx \sqrt{2m \Delta_*}/F_x$ and the absolute value of the mass at the Dirac points $\mathcal{M}\equiv |Z_\mu(t_0)|$: this allows one to easily check the formula $\Delta\varphi=-2\mu \arctan \left[\frac{\mathcal{D}}{\mathcal{M}}\frac{1+\chi \mu \tau}{2} \right]^\mu$.}\label{table5}
\vspace{0.25cm}
\begin{tabular}{c|c|c|c|c|c|c|c|c}
\#&$(\chi,\mu,\tau)$ &$X_{\chi,\tau}(t)$ &  $Y_{\chi,\tau}\,(t)$ & $Z_{\mu}(t)$ & $\Delta \varphi$ & $\mathcal{D}$&$\mathcal{M}$ &remarks \\
\hline
1&($-$, $+$, $+$) & $\f{F_x^2t^2}{2m}-\Delta_*$ & $c_y p_y$ & $M$ & $0$ &$c_y p_y$ &$M$&studied in \cite{FLM2012}\\
2&($-$, $+$, $-$) & $\f{F_x^2t^2}{2m}-\Delta_*$ & $c_y F_y t$ & $M$ & $-2 \arctan \biggl( \frac{c_y\sqrt{2m\Delta_*}}{M}\frac{F_y}{F_x}\biggr)$ &$c_y F_y\f{\sqrt{2m\Delta_*}}{F_x}$&$M$& mostly studied here  \\
3&($-$, $-$, $+$) & $\f{F_x^2t^2}{2m}-\Delta_*$ & $c_y p_y$ & $c_x F_x t$&  $2 \arctan \biggl( \frac{\sqrt{2m\Delta_*}}{p_y}\frac{c_x}{c_y}\biggr)$&$c_yp_y$&$c_x\sqrt{2m\Delta_*}$& \\
4&($-$, $-$, $-$) & $\f{F_x^2t^2}{2m}-\Delta_*$ & $c_yF_yt$ & $c_x F_x t$&$\pi$&$c_yF_y\f{\sqrt{2m\Delta_*}}{F_x}$&$c_x\sqrt{2m\Delta_*}$&\\
5&($+$, $+$, $+$) & $\f{F_x^2t^2}{2m}-\f{p_y^2}{2m}-\Delta_*$ & $\f{F_x p_y t }{m}$ & $M$& $-2 \arctan \biggl( \frac{p_y\sqrt{2m\Delta_*}}{m M}\biggr)$&$\frac{p_y\sqrt{2m\Delta_*}}{m}$&$M$& similar to \#2\\
6&($+$, $+$, $-$) & $\f{F_x^2t^2}{2m}-\f{F_y^2t^2}{2m}-\Delta_*$ & $\f{F_x F_y t^2}{m}$ & $M$&$0$&$\frac{F_y}{F_x}2\Delta_*$&$M$&similar to \#1\\
7&($+$, $-$, $+$) & $\f{F_x^2t^2}{2m}-\f{p_y^2}{2m}-\Delta_*$ & $\f{F_x p_y t}{m}$ & $c_x F_x t$&$\pi$&$\frac{p_y\sqrt{2m\Delta_*}}{m}$&$c_x\sqrt{2m\Delta_*}$&similar to \#4\\
8&($+$, $-$, $-$) & $\f{F_x^2t^2}{2m}-\f{F_y^2t^2}{2m}-\Delta_*$ & $\f{F_x F_y t^2}{m}$ & $c_x F_x t$& $2 \arctan \biggl( \frac{m c_x}{\sqrt{2m\Delta_*}}\frac{F_x}{F_y}\biggr)$&$\frac{F_y}{F_x}2\Delta_*$&$c_x\sqrt{2m\Delta_*}$&similar to \#3\\
\end{tabular}
\end{table*}
In this paper, we have mainly considered the special case of two cones with same mass, opposite chirality, traversed by a diagonal trajectory (case \#2). The case of a parallel trajectory (\# 1) was studied in \cite{FLM2012}. We do not elaborate here on the other cases which can be studied by similar techniques. Table~\ref{table5} presents our results for the phase shift $\Delta \varphi_{\chi,\mu,\tau}$ computed using the $N$-matrix approach for Landau-Zener crossings with complex gaps.  These results can be summarized as follows. Let $\mathcal{D}$ be the closest ``distance'' to the Dirac points (i.e. at the tunneling events when $t\approx \pm t_0$, $\mathcal{D}\equiv|Y_{\chi,\tau}(t_0)|$) and $\mathcal{M}$ be the absolute value of the mass at the Dirac points $\mathcal{M}\equiv |M_z(t_0)|=|Z_{\mu}(t_0)|$.  In all cases, we assumed that the tunneling events occur at $\pm t_0\approx \pm \sqrt{2m\Delta_*}/F_x$ corresponding to $X_{\chi,\tau}(\pm t_0)=0$.  

\begin{figure}
\begin{center}
\includegraphics[width=4cm]{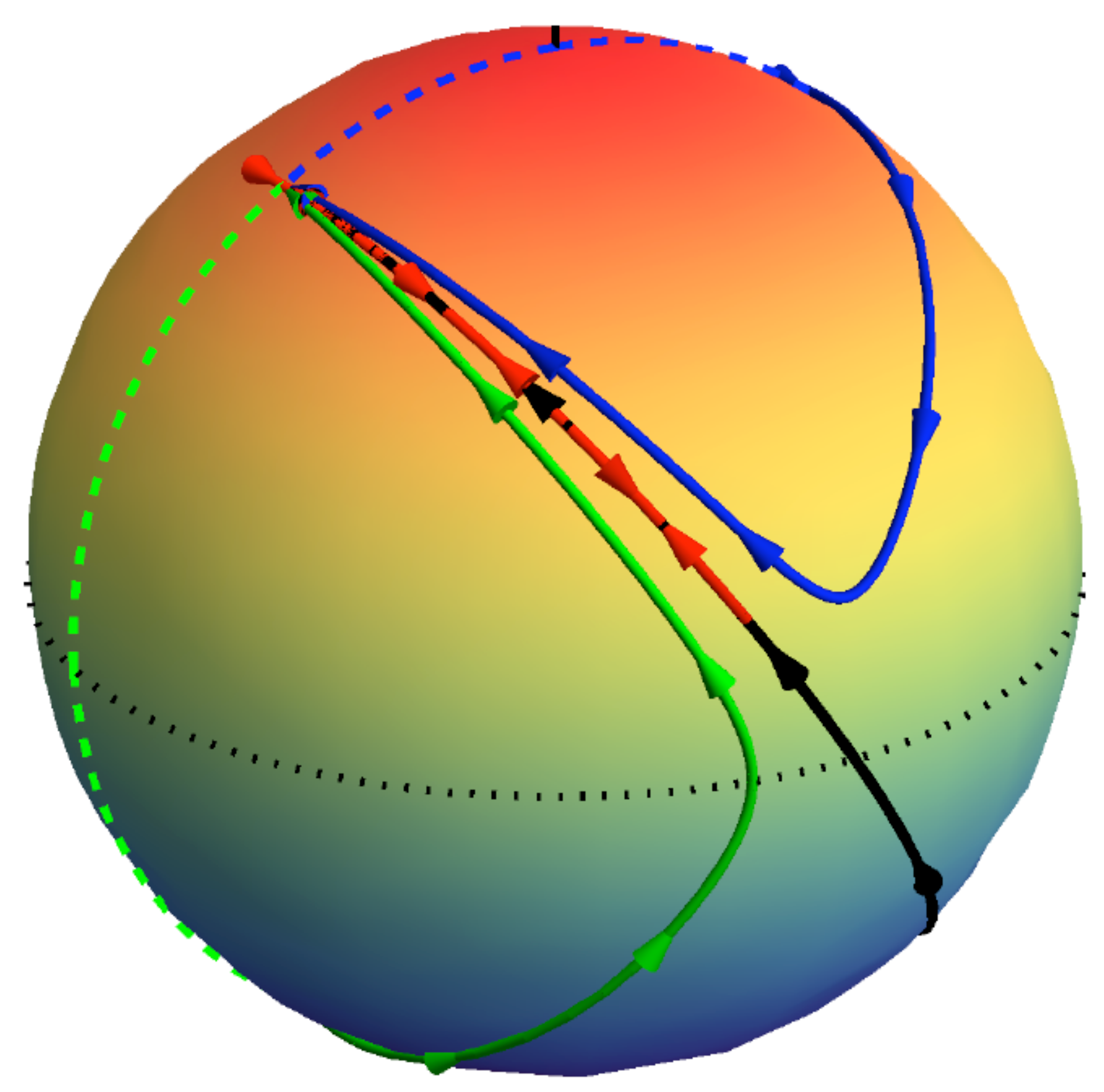}
\end{center}
\caption{Phase shift $\Delta \varphi$ as a solid angle on the Bloch sphere for the four cases in Table \ref{tablesmall}. Case \# 1 \& 6 is shown in red (null solid angle), case \# 2 \& 5 in blue (same as Fig. \ref{fig:geome}), case \#3 \& 8 in green and case \#4 \& 7 in black ($\pi$ solid angle, the geodesic closure is half of a great circle).}\label{fig:bs-4cases}
\end{figure}

It is possible to write a single formula for the eight cases. It is
\beq
\Delta\varphi=-2\mu \arctan \left[\frac{\mathcal{D}}{\mathcal{M}}\frac{1+\chi \mu \tau}{2} \right]^\mu
\eeq
where $\chi=\pm1$ is the product of the chirality of the two cones, $\mu=\pm1$ is the product of the mass sign of the two cones and $\tau=+1$ for parallel and $-1$ for diagonal trajectory. From the above formula, it is obvious that two different cases with the same $\mu$ and the same $\chi \tau$ have the same geometric phase shift. Actually, there are only four essentially different cases. In all cases $X(t)$ has the following structure $\sim t^2-\textrm{const.}=t^2-1$, $Y(t)$ either changes sign ($\chi \tau=+$) or does not ($\chi \tau=-$) between the two crossings, and similarly for $Z$, which either changes sign ($\mu=-$) or does not ($\mu=+$). This is summarized in Table \ref{tablesmall} and represented in Fig.~\ref{fig:bs-4cases}.
\begin{table}[h!]
\caption{Summary of phase shift in four essential cases.}\label{tablesmall}
\begin{center}
\begin{tabular}{c|c|c|c|c|c|c}
\#&$\chi\tau$ &$\mu$ & $X(t)$ & $Y(\mp t_0)$ & $Z(\mp t_0)$ & $\Delta \varphi$ \\
\hline
1 \& 6&-&+ & $t^2-1$ & $\mathcal{D}$ & $\mathcal{M}$ & 0\\
2 \& 5&+&+& $t^2-1$ & $\mp\mathcal{D}$ & $\mathcal{M}$ & $-2\arctan\frac{\mathcal{D}}{\mathcal{M}}$\\
3 \& 8&-&-& $t^2-1$ & $\mathcal{D}$ & $\mp\mathcal{M}$ & $2\arctan\frac{\mathcal{M}}{\mathcal{D}}$\\
4 \& 7&+&-& $t^2-1$ & $\mp\mathcal{D}$ & $\mp\mathcal{M}$ & $\pi$
\end{tabular}
\end{center}
\end{table}

In all cases, the validity of the \s regime is $\Delta_*\gg \sqrt{\mathcal{D}^2+\mathcal{M}^2}\geq \mathcal{D},\mathcal{M}$.

\section{Conclusions}\label{sec:conc}
The present paper extends our previous letter \cite{LFM2014}. It presents a careful derivation of the transition probability in a double cone St\"uckelberg interferometer. The Dirac cones are gapped and constitute avoided linear band crossings. We actually considered eight different cases having almost the same adiabatic energy spectrum but different adiabatic eigenstates. The differences come from the relative chirality (winding number) of the two Dirac cones, from their relative mass sign and from the reciprocal space trajectory crossing or not the line joining the two cones. We have shown that, in general, a \s interferometer does not only depend on the energy spectrum but is affected by geometrical properties revealing coupling between bands. We emphasize that the geometric phase revealed by \s interferometry depends both on diagonal (intra-band) and off-diagonal (inter-band) Berry connections.

Indeed, we generically found an additional phase shift of the \s interferences, which we relate to an open-path geometric phase involving the two bands. In the ``massless limit'' ($M_z\to 0$) where the spectrum becomes gapless (with two Dirac points), the open-path geometric phase becomes ambiguous. However, we showed there is an unambiguous $\pi$-shift in the \s interferences for case \# 2 (see also, for example, the toy model of a topological insulator at the beginning of \cite{LFM2014}). Further insight is gained from viewing the phase shift as the solid angle enclosed by the open-path on the Bloch sphere closed by the shortest geodesic. This is derived from older work of Pancharatnam in optical interferometry where he first realized the role of the polarization degree of freedom \cite{Pan1956}, and other works on open-path geometric phases \cite{SB1988,PS1998}. The geometrical area picture clarifies the role of the mass term in the Hamiltonian curve, and the observability of the open-path geometric phase.

\medskip

\acknowledgments
We thank Immanuel Bloch, Manuel Endres, Monika Schleier-Smith and Ulrich Schneider for useful discussions about their current experiment on St\"uckelberg interferometry in a honeycomb optical lattice.

\appendix

\section{Bloch oscillations for coupled bands with a scalar gauge potential}\label{sec:altder}
We give an alternative derivation of the equations (\ref{mainequations}). Instead of using a time-dependent vectorial gauge to include the effect of the external force (minimal coupling is $\hat{p}_x\to \hat{p}_x+F t$), we now use a scalar and time-independent gauge (minimal coupling is $\hat{H}\to \hat{H}+F \hat{x}$) and employ a well-known wavefunction ansatz due to Houston \cite{Houston} in the time-dependent Schr\"odinger equation.

For simplicity of notation, we consider a 1D crystal of lattice spacing $a$ caracterized by a Hamiltonian $\hat{H}$ describing the motion of electrons in the absence of an external force. This Hamiltonian is diagonalized by Bloch eigenstates $|\psi_{n,k}\rangle$ such that $\hat{H} |\psi_{n,k}\rangle=E_n(k)|\psi_{n,k}\rangle$, where $n$ is a band index and $k\in ]-\pi/a,\pi/a]$ a Bloch wavevector in the first Brillouin zone (BZ). One has $|\psi_{n,k}\rangle=e^{ik\hat{x}}|u_{n,k}\rangle$ where $\hat{x}$ is the position operator (i.e. the complete position operator, not just the position of the unit cell) and $|u_{n,k}\rangle$ is the cell-periodic part of the Bloch state such that $u_{n,k}(x+R)=u_{n,k}(x)$ for any Bravais lattice vector $R=a\times \textrm{integer}$. In the presence of a constant force $F$ the Hamiltonian becomes $\hat{H}-F\hat{x}$ (this is a time-independent scalar gauge choice). We now assume that the electron is initially in a Bloch state $|\psi(t=0)\rangle \equiv |\psi_{n_0,k_0}\rangle$ and want to solve its dynamics in the Bloch states basis $\{|\psi_{n,k}\rangle\}$. Expanding the state of the electron at time $t$ on this basis, we get $|\psi(t)\rangle=\sum_n \int_{BZ}dk c_{n,k}(t) |\psi_{n,k}\rangle$. We make an ansatz for $|\psi(t)\rangle$ following Houston \cite{Houston}:
\beq
|\psi(t)\rangle=\sum_n C_n(k(t))|\psi_{n,k(t)}\rangle \textrm{ with } k(t)=k_0+Ft
\eeq
where $C_n(k(t))$ are unknown expansion coefficients. This ansatz is inspired from our knowledge of Bloch oscillations and the equation describing the dynamics of electrons in crystals: $\frac{dk}{dt}=F$, which gives  $k(t)=k_0+Ft$. This ansatz is injected in the Schr\"odinger equation $i\frac{d}{dt}|\psi(t)\rangle = (\hat{H}-F\hat{x})|\psi(t)\rangle$. Next, we project on the Bloch state $|\psi_{n',k'(t)}\rangle$ to obtain:
\beqn
\delta(k'-k)i\frac{d}{dt}C_{n'}(k')&=&\delta(k'-k)E_{n'}(k')C_{n'}(k')\\
&-&F\sum_n C_n(k)\langle\psi_{n',k'}|(\hat{x}+i\partial_k)|\psi_{n,k}\rangle\nn
\eeqn
where we used that $\partial_t=F\partial_k$ (in the above equation we wrote $k$ for $k(t)=k_0+Ft$ to simplify the notations) and moved the term $\i\partial_k$ from the left to the right hand side. The matrix elements of the position operator in the Bloch state basis are \cite{LL9}:
\beq
\langle\psi_{n',k'}|\hat{x}|\psi_{n,k}\rangle=\delta_{n',n}i\delta'(k'-k)+\mathcal{A}_{n',n}(k')\delta(k'-k)
\eeq
where $\mathcal{A}_{n',n}(k)\equiv \langle u_{n',k}|i\partial_k|u_{n,k}\rangle$ is the Berry connection and $\delta'(x)$ is the derivative of the Dirac delta function (with $\delta'(-x)=-\delta'(x)$). Using the fact that $\langle\psi_{n',k'}|i\partial_k|\psi_{n,k}\rangle=i\partial_k[\delta_{n,n'}\delta(k-k')]=\delta_{n,n'}i\delta'(k-k')$, we find that $\langle\psi_{n',k'}|(\hat{x}+i\partial_k)|\psi_{n,k}\rangle=\mathcal{A}_{n',n}(k')\delta(k'-k)$. Eventually, the projected Schr\"odinger equation becomes:
\beq
i\frac{d}{dt}C_{n'}(k')=E_{n'}(k')C_{n'}(k')-F\sum_n C_n(k')\mathcal{A}_{n',n}(k')
\eeq
which we rewrite:
\beq
i\frac{d}{dk}C_{n}(k)=[\frac{E_{n}(k)}{F}-\mathcal{A}_{n}(k)]C_n(k)-\sum_{n'\neq n} \mathcal{A}_{n,n'}(k)C_n'(k)
\eeq
The first term in the right hand side gives the dynamical phase, the second the line integral of the diagonal Berry connection $\mathcal{A}_n(k)=\mathcal{A}_{n,n}(k)$ (these two terms together contribute to the adiabatically accumulated phase in the $n^{th}$ band) and the third term represents the coupling to other bands ($n'\neq n$) and depends on the off-diagonal Berry connection $\mathcal{A}_{n',n}(k)$.

In the particular case of only two bands $n=\pm$, if we define $A_n(t)\equiv C_n(k(t))e^{-i\int^t dt' E_n(k(t'))}e^{i\int^t dt' \mathcal{A}_n (t')}$, we obtain
\beqn
\frac{d}{dt}A_+&=&i\mathcal{A}_{+,-} A_- e^{i\int^t dt' [E_+ -E_-]}e^{i \int^t dt' [\mathcal{A}_- -\mathcal{A}_+]}\nn\\
\frac{d}{dt}A_-&=&i\mathcal{A}_{-,+} A_+ e^{i\int^t dt' [E_- -E_+]}e^{-i \int^t dt' [\mathcal{A}_+ -\mathcal{A}_-]}\nn
\eeqn
where $\mathcal{A}_{n,n'}(t)\equiv \langle u_{n,k(t)}|i\frac{d}{dt}|u_{n',k(t)}\rangle=F \langle u_{n,k(t)}|i\partial_k |u_{n',k(t)}\rangle=F\mathcal{A}_{n,n'}(k(t))$. These are exactly the equations (\ref{mainequations}) upon using $E_-=-E_+$ and $\mathcal{A}_{n,n'}(t)=\langle \psi_n(t)|i\partial_t|\psi_{n'}(t)\rangle$, i.e. $|\psi_n(t)\rangle=|u_{n,k(t)}\rangle$.

\begin{table}
\caption{Summary of the coordinate parameterizations and their time derivatives for the Hamiltonian written in the graphene (gr) and Landau-Zener (lz) bases, respectively, with the adiabatic spectrum $E_+(t)=[(t^2-\Delta_*)^2+c_y^2 F_y^2 t^2+M^2]^{1/2}$.}\label{tableappendix}
\vspace{0.25cm}
\begin{tabular}{c|c|c}
  & graphene basis & Landau-Zener basis \\
\hline
& & \\
$B_x$ & $t^2-\Delta_*$  & $c_y F_y$  \\
& & \\
$B_y$ & $c_y F_y$ &  $M$\\
 & & \\
$B_z$ &  $M$ &  $t^2-\Delta_*$\\
& & \\
$\sin\theta$ & $\frac{\sqrt{(t^2-\Delta_*)^2+c_y^2 F_y^2 t^2}}{E_+}$  & $\frac{\sqrt{c_y^2F_y^2t^2+M^2}}{E_+}$ \\
& & \\
$\cos\theta$  & $\frac{M}{E_+}$  &   $\frac{t^2-\Delta_*}{E_+}$  \\
& & \\
$e^{i\phi}$&$\frac{t^2-\Delta_*+ic_y F_y t}{\sqrt{(t^2-\Delta_*)^2+c_y^2 F_y^2 t^2}}$& $\frac{c_yF_yt+iM}{\sqrt{c_y^2F_y^2t^2+M^2}}$\\
& & \\
$\theta (t_{i,f})$ & $\arcsin (\f{c_y F_y \sqrt{\Delta_*}}{\sqrt{c_y^2 F_y^2 \Delta_*+M^2}})$ & $\f{\pi}{2}$\\
& & \\
$\phi(t_f)-\phi(t_i)$ & $-\pi$ & $-2\arctan\biggl(\f{c_y F_y\sqrt{\Delta_*}}{M} \biggr)\equiv\Delta\varphi$  \\
& & \\
$\phi(0)$ & $\pi$ & $\f{\pi}{2}$\\
& & \\
$\dot{\theta}$& $\frac{M t [2(t^2-\Delta_*)+c_y^2 F_y^2 ]}{E_+^2 \sqrt{(t^2-\Delta_*)^2+c_y^2F_y^2t^2}}$ &$-\frac{t[c_y^2F_y^2(t^2+\Delta_*)+2M^2]}{E_+^2 \sqrt{c_y^2F_y^2t^2+M^2}}$\\
& & \\
$\dot{\phi}$&$-\frac{c_yF_y(t^2+\Delta_*)}{(t^2-\Delta_*)^2+c_y^2F_y^2t^2}$ &$-\frac{c_yF_yM}{c_y^2F_y^2t^2+M^2}$\\
& & \\
\hline
\end{tabular}
\end{table}

The understanding gained from this alternative derivation of the main equations is threefold: (i) These equations are exact and do not rely on a classical treatment of the orbital motion and a quantum treatment of the internal state (band indices) dynamics. Indeed $\frac{dk}{dt}=F$ is actually exact (see, for example, page 1971 of Ref. \cite{Xiao2010}). These equations are not restricted to a semiclassical regime and can be used whatever the magnitude of the force (from the adiabatic to the sudden regime). (ii) The matrix elements of the position operator in the Bloch states are crucial. It is the complete position operator $\hat{x}$ appearing. In the case of a lattice with several sites in the unit cell (such as the dimerized chain with two sublattices $A$ and $B$), this automatically selects a Bloch Hamiltonian $\hat{H}(k)=e^{-ik\hat{x}}\hat{H}e^{ik\hat{x}}$ written in the so-called basis II rather than $\hat{H}_I(k)=e^{-ik\hat{R}}\hat{H}e^{ik\hat{R}}$ written in the so-called basis I, where $\hat{R}$ is the position operator of the unit cell only and not the full position operator $\hat{x}=\hat{R}+\hat{\delta}$, where $\hat{\delta}$ gives the relative position within the unit cell. Note that $\hat{H}(k)$ is not periodic in $k$ with the BZ periodicity (in contrast to $\hat{H}_I(k))$. Basis I versus basis II issues are discussed in \cite{BM2009,Fuchs2010,Fruchart2014}. It is important to pay attention to that when discussing motion in reciprocal space that crosses the edge of the BZ. (iii) The force can be included in whatever gauge. We showed derivation in both the time-dependent vectorial gauge ($H(\hat{p}_x)\to H(\hat{p}_x+Ft)$) (for more details in a similar case see \cite{FLM2012}) and the time-independent scalar gauge ($\hat{H}=H(\hat{p}_x)\to \hat{H}-F\hat{x}$).

\section{Standard parametrization for a spin-1/2 Zeeman Hamiltonian}\label{sec:appenzeeman}
Given the $2\times2$ Hamiltonian
\beqn
H(t)=\vec{B}(t)\cdot \vec{\sigma}=E_+(t)\left(\begin{array}{cc}\cos \theta&\sin \theta e^{-i \phi}\\\sin \theta e^{i \phi}&-\cos \theta\end{array} \right),
\eeqn
where $\vec{B}(t)=(B_x,B_y,B_z)$ and $E_+\equiv |\vec{B}|$ (here and the following, the time-dependence of the parameters are assumed) we have
\beq
\sin\theta=\f{\sqrt{B_x^2+B_y^2}}{E_+},\tr{\ \ }\cos\theta=\f{B_z}{E_+},\eeq
and
\beq
e^{i\phi}=\f{B_x+iB_y}{\sqrt{B_x^2+B_y^2}}.
\eeq
Accordingly, their time derivatives are given by
\beqn
&&\dot{\theta}=\f{B_z(B_x \dot{B}_x +B_y \dot{B}_y +B_z \dot{B}_z)-E_+^2 \dot{B}_z}{E_+^2\sqrt{B_x^2+B_y^2}}\\
&&\dot{\phi}=\f{\dot{B}_x B_y-\dot{B}_y B_x}{B_x^2+B_y^2}
\eeqn
For the two main Hamiltonians we study in the paper (case \#2 written in two Hamiltonian bases), we summarize the parameterizations and their derivatives in Table~\ref{tableappendix}.

\section{First order adiabatic perturbation theory: integral in the complex plane}\label{sec:appintcomplex}
\subsubsection{Analytic structure of the integral}\label{sec:sec:ana}
\begin{figure}
\begin{center}
\includegraphics[width=4.5cm]{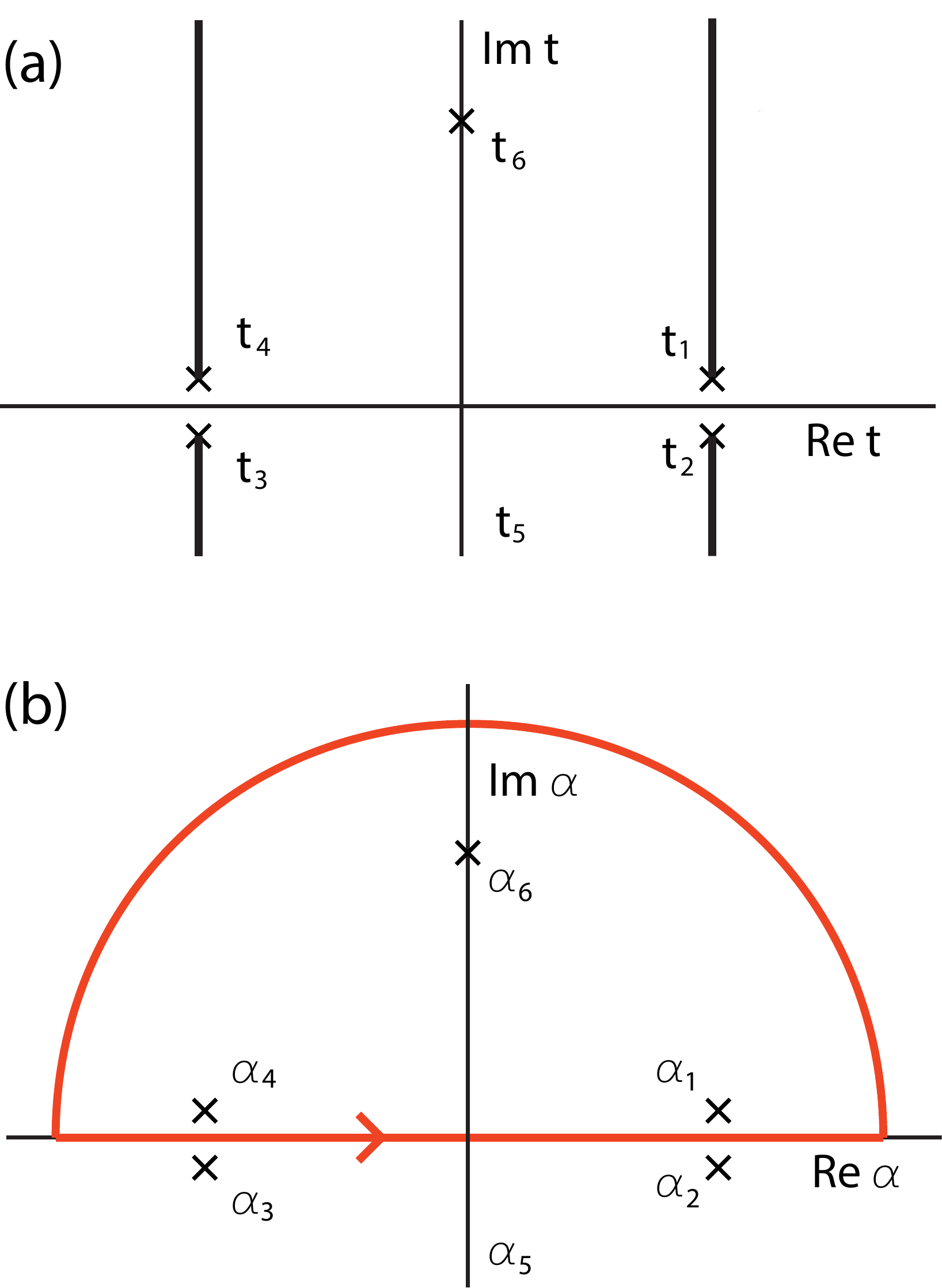}
\end{center}
\caption{(a) The analytic structure of the function $g(t)\dot{\alpha}e^{i\alpha(t)}$ in the complex $t$ plane. The function has four branch points (because of $\alpha(t)$) in $t_1$ to $t_4$. In addition there are five poles at $t_1,t_2,t_3,t_4$ and $t_6$. (b) In complex $\alpha$ plane, the function $g(\alpha)e^{i\alpha}$ has five simple poles ($\alpha_1$, $\alpha_2$, $\alpha_3$, $\alpha_4$ and $\alpha_6$) indicated by crosses, three of which are in the upper complex plane ($\alpha_1$, $\alpha_4$ and $\alpha_6$). The integration contour is shown in red. The point $\alpha_5=-\alpha_6$ is also indicated but is not a pole of $g(\alpha)$.}
\label{branchcut}
\end{figure}

It is known that a direct complex time integration of
\beqn
A_+(+\infty)&=& \int_{-\infty}^{\infty} dt  \f{\dot{\theta}-i\dot{\phi}\sin \theta}{2}\,e^{i\beta(t)},
\eeqn
obtained generally in the adiabatic perturbation theory \cite{LLbook}, is inconvenient due to the presence of branch cuts, see Fig.~\ref{branchcut}(a). Following Ref.~\cite{BM1972} we make a change of integration variable from the time variable $t$ to the phase variable given by the dynamical phase $\alpha(t)\equiv \int_0^t dt' 2E_+(t')$ so that $\int dt=\int d\alpha (1/\dot{\alpha})$ \cite{foot1}. For the case of Dirac cones with a constant mass, an opposite chirality and a diagonal trajectory (case \# 2) in the Landau-Zener basis, we have $A_+(+\infty)=\int_{-\infty}^{\infty} d\alpha f(\alpha)$ with
\begin{widetext}
\beqn
f(\alpha)
=-i\frac{-c_y^2F_y^2t(t^2+\Delta_*)-2tM^2+ic_yF_y M E_+(t)}{4\sqrt{c_y^2F_y^2t^2+M^2}E_+(t)^3}e^{i\int_0^t dt'\frac{c_yF_yM}{c_y^2F_y^2t'^2+M^2}\frac{t'^2-\Delta_*}{E_+(t')}+i\alpha},
\label{gofalpha}
\eeqn
\end{widetext}
where we use the fact that $-\phi(t)+\int_0^t dt' \dot{\phi}(1-\cos\theta)=-\int_0^t dt' \dot{\phi}\cos\theta-\phi(0)$ with $\phi_{\tr{lz}}(0)=\pi/2$.

------------------

To see the merit of changing integration variable, we now analyze the structure of $f(\alpha)$.
In the original complex $t$-plane, there are only 4 branch points related to the function of the dynamical phase $\exp[i\int^t_0 dt' 2 E_+(t')]$
emanating from the 4 poles of $f(\alpha) \dot{\alpha} e^{-i\alpha}$ (before the change of variable) when $E_+(t)=0$, giving
\beqn
&&t_1\approx \sqrt{\Delta_*}+ic_y F_y/2,\tr{\ \ \ }t_2=t_1^*,\nn\\
&&t_3=-t_1 \tr{\ \ \ and\ \ \ } t_4=-t_1^*.
\eeqn
In the $\alpha$-plane with $\alpha_j\equiv \alpha(t_j)$, we define the complex line element as
\beq
\int_0^{t_j}dt'(\ldots)= \int_0^{\tr{Re} t_j}du (\ldots) + i \int_0^{\tr{Im} t_j}dv (\ldots)|_{t'=\tr{Re} t_j+i v}\eeq
with $t'\equiv u+ i v$ and $u,v$ being real. The location corresponding to the $t_j$'s are then given by $\alpha_1 \approx 4\Delta_*^{3/2}/3+i\pi c_y^2 F_y^2 \sqrt{\Delta_*}/4$, $\alpha_2=\alpha_1^*$, $\alpha_3=-\alpha_1$ and $\alpha_4=-\alpha_1^*$. These are computed in the \s limit, e.g., $\textrm{Im}\alpha_1=\textrm{Re}\int_0^{\textrm{Im}t_1}dv 2 E_+(\textrm{Re}t_1+iv)= \textrm{Re}\int_0^{\textrm{Im}t_1}dv 2 \sqrt{-4\Delta_*v^2+c_y^2F_y^2(\Delta_*+2i\sqrt{\Delta_*}v)+M^2}\approx \int_0^{\textrm{Im}t_1}dv 2 \sqrt{\Delta_*(c_y^2F_y^2-4v^2)}=\pi c_y^2F_y^2\sqrt{\Delta_*}/4$.

The behavior of the function $f(\alpha)$ around $\alpha_j$ is determined as follow. First, we compute the change of $\alpha$ around $\alpha_1$, in terms of the variable $t$ around $t_1$, by making an expansion $\alpha(t)=\int_0^t dt' 2E_+(t')$ around $t=t_1$. This gives $\alpha-\alpha_1\approx 8\sqrt{i\Delta_* c_y F_y}(t-t_1)^{3/2}/3$. Next, by
making an expansion of $f(\alpha(t))$ around $t_1$ and using the former relation, we get to the leading order in $\alpha_1$ the expression
\beqn\label{res1}
f(\alpha)\approx
\biggl(\f{1}{6}e^{-i\int_0^{t_1} dt' \dot{\phi_{\tr{lz}}}\cos\theta_{\tr{lz}}+i\alpha_1} \biggr)\f{1}{\alpha-\alpha_1},
\eeqn
which shows that $\alpha_1$ is a pole in the $\alpha$ plane (rather than a branch point as $t_1$).

Second, an expansion around $t= t_4$ for $\alpha(t)$ similarly yields $\alpha-\alpha_4\approx 8\sqrt{i\Delta_* c_y F_y}(t-t_4)^{3/2}/3$. For the expansion of $f(\alpha(t))$ we have \beqn
f(\alpha)\approx\label{res4}
\biggl(-\f{1}{6}e^{-i\int_0^{t_4} dt' \dot{\phi_{\tr{lz}}}\cos\theta_{\tr{lz}}+i\alpha_4} \biggr)\f{1}{\alpha-\alpha_4}
\eeqn
showing that it is again a simple pole with an additional minus sign. These expressions also give the residues around $\alpha_{1,4}$, which we will use in the next section when computing the contour integral. Similar structures are obtained for $\alpha\approx \alpha_{2,3}$ as simple poles.

From the expression of Eq. (\ref{gofalpha}), there seems to be two additional branch points (both in $t$- and in $\alpha$-planes) coming from $\sqrt{c_y^2F_y^2t^2+M^2}=0$ which are located at:
$t_6=iM/(c_yF_y)$ and $t_5=-t_6$  or in other words at $\alpha_6=2\int_0^{t_6}E_+ (t) \approx 2i \Delta_* M/c_y F_y$ and $\alpha_5=-\alpha_6$,
A little thinking
actually shows that these are not branch points in the complex $\alpha$-plane but a simple pole at $\alpha_6$ and no pole at $\alpha_5$.
To see this,
we let $t=iy$ on the imaginary axis with real $y$, then $\cos \theta_{\tr{lz}}\approx -(y^2+\Delta_*)/\sqrt{(y^2+\Delta_*)^2}=-1$ when $\Delta_*$ is large. The phase factor is then approximately $e^{-i\int_0^t dt' \dot{\phi_{\tr{lz}}}\cos\theta_{\tr{lz}}}\approx e^{i\int_0^t dt' \dot{\phi_{\tr{lz}}}}=-i e^{i\phi_{\tr{lz}}(t)}$.
We arrive, in a rough approximation, at
\beqn
f(\alpha)&\approx& -\frac{\dot{\theta_{\tr{lz}}}-i\dot{\phi_{\tr{lz}}}\sin \theta_{\tr{lz}}}{2\dot{\alpha}}e^{i\phi_{\tr{lz}}+i\alpha}\nn\\&=&\frac{c_y^2F_y^2t(t^2+\Delta_*)+2tM^2-ic_yF_y M E_+(t)}{4E_+(t)^3(c_yF_yt-iM)}e^{i\alpha}
\label{grough}
\eeqn
on the imaginary axis. We thus see that the apparent poles at $t_5$ and $t_6$ reduce to a single pole at $t_6$. By making an expansion around $\alpha_6$ we get
\beq\label{res6}
f(\alpha)\approx i\frac{M^3\, e^{i\alpha_6}}{2\Delta_*^2c_y^2F_y^2}\,\f{1}{\alpha-\alpha_6},
\eeq
which shows that it is also a simple pole in the $\alpha$ plane.

In summary, the function $f(\alpha)$ displays 5 simple poles located at
\beqn\label{6poles}
&&\alpha_1\approx \f{4}{3}\Delta_*^{3/2}+\f{i\pi}{4} c_y^2 F_y^2 \sqrt{\Delta_*}, \tr{\ }\alpha_2=\alpha_1^*,\tr{\ }
\alpha_3=-\alpha_1,\nn\\&&\alpha_4=-\alpha_1^*, \tr{\ \ }\alpha_6\approx 2i\Delta_* M/(c_yF_y),
\eeqn
see Fig. \ref{branchcut}. When studying the function $f(\alpha)$ in the vicinity of the imaginary axis, we shall use (\ref{grough}), elsewhere we will use the complete expression (\ref{gofalpha}).

\subsubsection{Residues of $A_+(\infty)$}\label{sec:sec:res}
To evaluate the expression $A_+(\infty)$ we use the red contour shown in Fig. \ref{branchcut}(b) and the residue theorem gives
\beqn
&&A_+(\infty)+\int_\textrm{UHP}d\alpha f(\alpha)=\oint d\alpha f(\alpha)\nn\\
&&\tr{\ }=2\pi i [\textrm{Res}(f,\alpha_1)+\textrm{Res}(f,\alpha_4)+\textrm{Res}(f,\alpha_6)].
\eeqn
The line integral over the upper half plane (UHP) vanishes exponentially thanks to the factor $e^{i\alpha}$ when $\textrm{Im} \alpha>0$. Therefore $A_+(\infty)=2\pi i \sum_{j=1,4,6}\textrm{Res}(f,\alpha_j)$. The residues for the first two simple poles are given as $\textrm{Res}(f,\alpha_j)=f(\alpha) (\alpha-\alpha_j)|_{\alpha=\alpha_j}$ for $j=1,4$, using Eqs. (\ref{res1}) and (\ref{res4}).

Next, we consider the residue at $\alpha_6$. From the fact that $\alpha_6$ is further away from the real axis than $\alpha_1$ and $\alpha_4$, we already see that its contribution to $A_+$ will be negligible. Using Eqs. (\ref{res6}), (\ref{6poles}), we have $\textrm{Res}(f,\alpha_6)\approx i\frac{M^3}{2\Delta_*^2c_y^2F_y^2}e^{-2\Delta_*M/c_yF_y}$. Indeed $e^{-\textrm{Im}\alpha_6}\ll e^{-\textrm{Im}\alpha_1}$ when $\Delta_*$ is large, which is reminiscent of the Landau argument for tunneling in the adiabatic limit \cite{LLbook}. Furthermore, note that the residue vanishes also at $M\to 0$.

The sum of the two dominant residues becomes
\beqn\label{res}
&&\textrm{Res}(f,\alpha_1)+\textrm{Res}(f,\alpha_4)=\frac{i}{6}(e^{i\beta_1}-e^{i\beta_4})\nn\\
&&=\frac{i}{6}(e^{- \textrm{Im}\beta_1+i \textrm{Re}\beta_1}-e^{-\textrm{Im}\beta_4+i \textrm{Re}\beta_4}).
\eeqn
where we define $\beta_{1,4}\equiv\beta(t_{1,4})$ and we recall that $\beta(t)=-\pi/2-\int_0^t dt'\dot{\phi_{\tr{lz}}}\cos\theta_{\tr{lz}}+\alpha(t)$.

\section{Integral giving the geometric phase in the graphene basis}\label{sec:appenx}
The integral $I=\int_{-\sqrt{\Delta_*}}^{\sqrt{\Delta_*}}dt \dot{\phi_{\tr{gr}}}\cos\theta_{\tr{gr}}$ in the graphene basis is given by
\begin{widetext}
\beqn
I=-2c_y F_y M \int_0^{\sqrt{\Delta_*}} dt \frac{t^2+\Delta_*}{(t^2-\Delta_* )^2+c_y^2F_y^2 t^2}\frac{1}{\sqrt{(t^2-\Delta_*)^2+c_y^2F_y^2 t^2+M^2}}
\eeqn
By perfecting the square for the variable $t$ and using that $\Delta_*\gg c_y^2 F_y^2$ we get
\beqn
I\approx -2c_yF_yM \int_0^{\sqrt{\Delta_*}} dt \frac{t^2+\Delta_*}{(t^2-\Delta_*)^2+c_y^2 F_y^2 \Delta_*}\frac{1}{\sqrt{(t^2-\Delta_*)^2+c_y^2F_y^2 \Delta_*+M^2}}.
\eeqn
One notices that the main contribution of the integral comes from $t\approx \sqrt{\Delta_*}$ as $c_y^2 F_y^2 \rightarrow 0$. So we expand around the value $t=\sqrt{\Delta_*}-\epsilon$ and keeping the leading contribution of $\epsilon$:
\beqn
I\approx -2c_yF_yM\int_0^{\sqrt{\Delta_*}} d\epsilon \frac{2\Delta_*}{4 \Delta_*\epsilon^2+c_y^2F_y^2\Delta_*}\frac{1}{\sqrt{4\Delta_*\epsilon^2+c_y^2F_y^2\Delta_*+M^2}}.
\eeqn
The integral can be evaluated using the identity
\beqn
\int\frac{d x}{(x^2+a)\sqrt{x^2+b}}=\frac{1}{\sqrt{a b-a^2}}\arctan\biggl( \frac{x \sqrt{b-a}}{\sqrt{a x^2+a b}} \biggr)
\eeqn
to give
\beqn
I\approx - 2 \arctan\frac{M}{c_yF_y\sqrt{\Delta_*}}=2 \arctan\frac{c_yF_y\sqrt{\Delta_*}}{M}-\pi.
\eeqn
\end{widetext}

\section{Adiabatic perturbation theory for the massless case}\label{sec:appintcomplex2}
\begin{figure}[ht]
\begin{center}
\includegraphics[width=5cm]{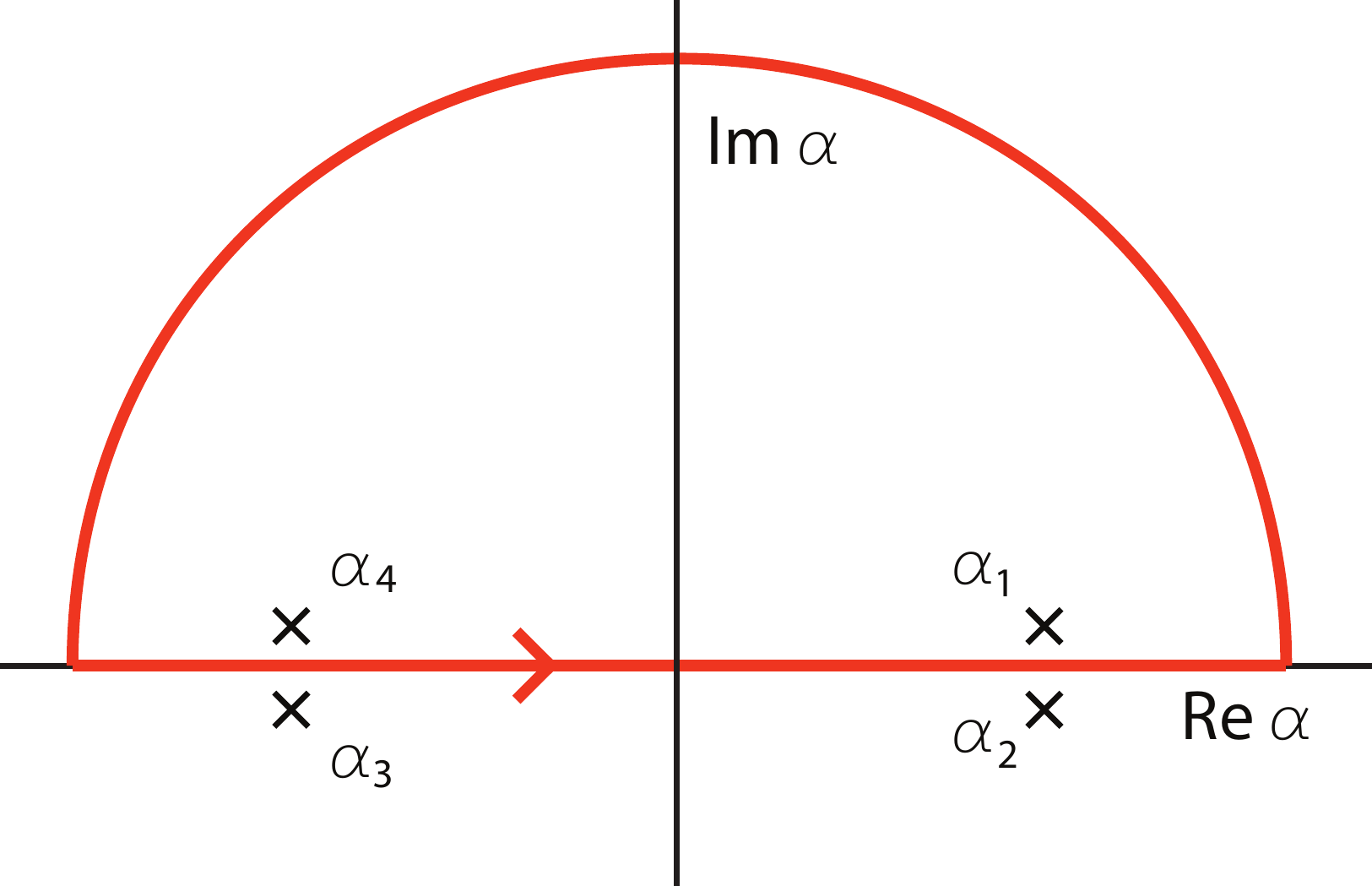}
\end{center}
\caption{The analytic structure of $g(\alpha)e^{i\alpha}$ for the massless case. The function has 4 simple poles ($\alpha_1$, $\alpha_2$, $\alpha_3$ and $\alpha_4$) indicated by crosses, two of which are in the upper complex plane ($\alpha_1$, $\alpha_4$). The integration contour is shown in red.}
\label{branchcut2}
\end{figure}

The full expression for the amplitude for case \#2 with $M=0$, in the graphene basis and south pole gauge, is given by $A_+(\infty)=\int_{-\infty}^{\infty} d\alpha g(\alpha)e^{i\alpha}$ with:
\beq
g(\alpha)=-i\frac{c_yF_y(t^2+\Delta_*)}{4E_+^3}
\label{gofalpha2}
\eeq
and $E_+(t)=\sqrt{(t^2-\Delta_*)^2 +c_y^2 F_y^2 t^2}$. There are only 4 poles in $\alpha$ (coming from $E_+^3=0$ and located at $\alpha_1\approx4\Delta_*^{3/2}/3+i\frac{\pi}{4}c_y^2F_y^2\sqrt{\Delta_*}$, $\alpha_2\approx\alpha_1^*$, $\alpha_3\approx -\alpha_1$ and $\alpha_4\approx -\alpha_1^*$) and no branch cuts, see Fig.~\ref{branchcut2}. The contour is therefore as before but there are only two residues to compute.

In both cases ($j=1$ and $4$), the relationship between the change in $\alpha$ around the pole $\alpha_j$ and $t$ around the branch point $t_j$ is obtained similarly by making an expansion around $t=t_j$ for the function $\alpha(t)=\int_0^t dt' 2E_+(t')$ giving $\alpha-\alpha_j\approx 8\sqrt{ic_yF_y\Delta_*}(t-t_j)^{3/2}/3$. Plugging this into the leading order expansion of $g(\alpha)$ around the poles and using the fact that $E_+(t\sim t_j)^3\sim 3ic_yF_y\Delta_*(\alpha-\alpha_j)$, we find $g(\alpha\sim \alpha_1)\approx -\frac{1}{6(\alpha-\alpha_1)}$ and $g(\alpha\sim\alpha_4)\approx -\frac{1}{6(\alpha-\alpha_4)}$. We thus see that both poles give the same residues, same as for the massive case in the graphene basis, see section \ref{hamgrbasis}.

\section{Proof that the open-path geometric phase equals half of a solid angle}\label{sec:geodrule}
\begin{figure}[ht]
\begin{center}
\includegraphics[width=3.6cm]{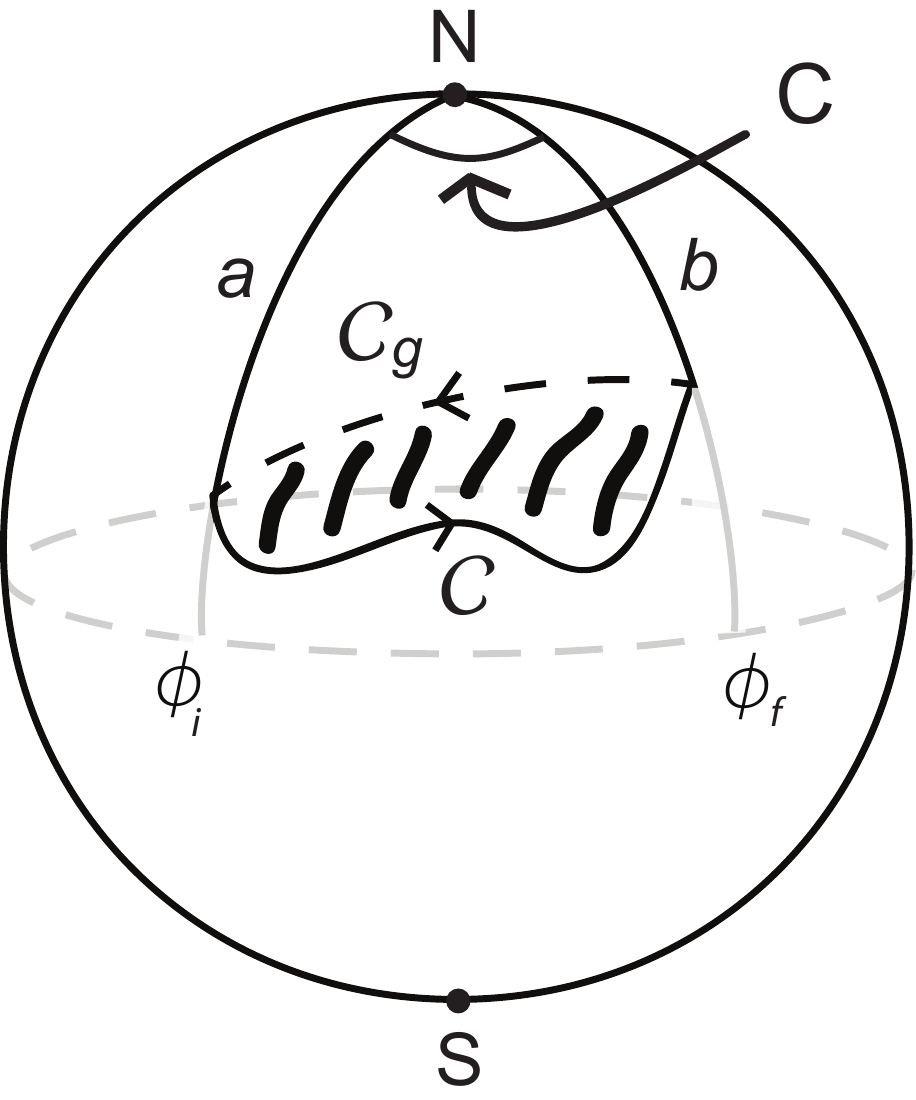}
\end{center}
\caption{The shaded region area is bounded by the two paths $\mathcal{C}$ and $\mathcal{C}_g$, which are defined by the Hamiltonian trajectory $\mathcal{C}$ and the shortest geodesic $\mathcal{C}_g$ connecting the final and initial points.}\label{fig:geo}
\end{figure}

In this appendix, we give an elementary proof that the open-path geometric phase $\Gamma\equiv \int_\mathcal{C} \langle \psi| id |\psi\rangle+\textrm{arg}\langle \psi_i|\psi_f\rangle=\int_{t_i}^{t_f} dt \langle \psi(t)| i\partial_t |\psi(t)\rangle+\textrm{arg}\langle \psi(t_i)|\psi(t_f)\rangle$ for a spinor (spin 1/2) $|\psi\rangle$ is given by half of the solid angle (or half of the area) of the open trajectory closed by the shortest geodesic (which is a portion of great circle in the case of a sphere). The open path $\mathcal{C}$ on the Bloch sphere (see Fig.~\ref{fig:geo}) is parametrized by $t$ going from $t_i$ to $t_f$ so that the state $|\psi(t)\rangle$ depends on $t$. Because this phase is gauge independent (see section \ref{sec:twobands}), we can choose a specific gauge to compute it. We therefore parametrize the spinor by $|\psi\rangle = \left(\begin{array}{c}\cos(\theta/2)\\\sin(\theta/2)e^{i\phi}\end{array} \right)$, which is well defined except at the south pole ($\theta=\pi$) where it suffers a phase ambiguity. In the following, we first compute the line integral of the Berry connection and then the argument of the scalar product.

The line integral of the Berry connection is $\int_\mathcal{C} \langle \psi| id |\psi\rangle=-\int_{\phi_i}^{\phi_f} d\phi \sin^2(\phi/2)$. This is actually equal to (minus) half of the solid angle $\Omega_1$ of a closed loop constructed from the open path $\mathcal{C}$ and the two meridians (called a and b in Fig.~\ref{fig:geo}) relating the north pole ($\theta=0$) either to ($\theta_i,\phi_i$) or to ($\theta_f,\phi_f$). Considering $\theta$ as a function of $\phi$, this solid angle is $\Omega_1=\int_{\phi_i}^{\phi_f} d\phi\int_0^{\theta(\phi)}d\theta\sin\theta=2 \int_{\phi_i}^{\phi_f} d\phi \sin^2(\phi/2)$ as the area measure on the sphere is $d\phi d\theta \sin \theta$. Therefore $\int_\mathcal{C} \langle \psi| id |\psi\rangle=-\Omega_1/2$.

The overlap between the initial and final spinors is given by
\beq
\langle \psi_i|\psi_f\rangle=\cos\frac{\theta_i}{2}\cos\frac{\theta_f}{2}+\sin\frac{\theta_i}{2}\sin\frac{\theta_f}{2}e^{i(\phi_f-\phi_i)}
\eeq
We call $\Phi\equiv \textrm{arg}\langle \psi_i|\psi_f\rangle $. Then
\beq
\tan \Phi = \frac{\sin\frac{\theta_i}{2}\sin\frac{\theta_f}{2}\sin(\phi_f-\phi_i)}{\cos\frac{\theta_i}{2}\cos\frac{\theta_f}{2}+\sin\frac{\theta_i}{2}\sin\frac{\theta_f}{2}\cos(\phi_f-\phi_i)}
\eeq
which, from elementary spherical geometry \cite{Todhunter1886}, we recognize as equal to $\tan \Omega_2/2$ where $\Omega_2$ is the solid angle of the triangle on the sphere (also known as the spherical excess $E$) with corners located at the north pole ($\theta=0$), at the initial point ($\theta_i,\phi_i$) and at the final point ($\theta_f,\phi_f$). This triangle is made of three portions of great circles (among which two meridians of length $a=\theta_i$ and $b=\theta_f$; their included angle being $C=\phi_f-\phi_i$). Therefore $\Phi=\textrm{arg}\langle \psi_i|\psi_f\rangle=\Omega_2/2$.

Combining this two partial results, we obtain that the open-path geometric phase $\Gamma=(-\Omega_1+\Omega_2)/2=-\Omega/2$, where $\Omega\equiv \Omega_1-\Omega_2$ is the area of the loop made of the open path $\mathcal{C}$ closed by the shortest geodesic (portion of great cicrle) $\mathcal{C}_g$ going from ($\theta_f,\phi_f$) to ($\theta_i,\phi_i$).

\begin{thebibliography}{99}
\bibitem{footnoteblochhamil}It is important to realize that we are here dealing with Bloch Hamiltonians $H(\vec{k})=e^{-i\vec{k}\cdot \vec{r}}He^{i\vec{k}\cdot \vec{r}}$, where $\vec{r}$ is the position operator, and not true Hamiltonians $H$. Bloch Hamiltonians have the cell-periodic part $|u_{n,\vec{k}}\rangle$ of the Bloch wavefunctions $|\psi_{n,\vec{k}}\rangle=e^{i\vec{k}\cdot \vec{r}} |u_{n,\vec{k}}\rangle$ as eigenstates: $H(\vec{k})|u_{n,\vec{k}}\rangle= E_n(\vec{k})|u_{n,\vec{k}}\rangle$. Because Bloch Hamiltonians depend on a continuous parameter (the wavevector $\vec{k}$), it is possible to apply the Berry construction to them. In our work, the $|\psi_\pm(t)\rangle$, which are adiabatic eigenstates of $H(t)\equiv H(\vec{k} \to \vec{F}t)$ are therefore descendants of the $|u_{n=\pm,\vec{k}}\rangle$ with $\vec{k}$ replaced by $\vec{F}t$. See also appendix \ref{sec:altder}.
\bibitem{foot1} To get a feeling for the change of variable from $t$ to $\alpha$, we consider $t$ real and see that $\alpha$ is then also real. In addition, $\dot{\alpha}=2E_+>0$ so that $\alpha$ is an increasing function of $t$. In two limits, we obtain $\alpha \sim 2\Delta_* t$ (when $t^2\ll \Delta_*$) and $\alpha \sim 2 t^3/3$ (when $t^2\gg \Delta_*$). So that $\alpha$ covers the whole real line when $t$ spans the reals.
\end{thebibliography}
\end{document}